\documentstyle[12pt]{article}
\def\lesssim{\mathrel{\hbox{\rlap{\hbox{\lower4pt\hbox{$\sim$}}}\hbox{\raise1pt\hbox{$<$}}}}}
\def\gtrsim{\mathrel{\hbox{\rlap{\hbox{\lower4pt\hbox{$\sim$}}}\hbox{\raise1pt\hbox{$>$}}}}}
\newcommand{\Tc}{T_{c}}
\newcommand{\Tu}{T_{u}}
\newcommand{\Tb}{T_{b}}

\newcommand{\To}{T_{0}}
\newcommand{\Tn}{T_{n}}

\begin{document}

\title{The Effects of Electro-weak Phase Transition Dynamics on
  Baryogenesis and Primordial Nucleosynthesis} \author{Andrew F.
  Heckler$^{1}$ \\University of
    Washington\\ FM-15 Seattle, Washington} \maketitle
\begin{abstract}
  The evolution of the electro-weak phase transition, including
  reheating due to the release of latent heat in shock waves, is
  calculated for various values of as yet unknown parameters of
  electro-weak theory such as latent heat and bubble wall surface
  tension. We show that baryon production, which occurs in the
  vicinity of the bubble walls of the phase transition, can be a
  sensitive function of bubble wall velocity, and this velocity
  dependence is important to include in the calculation of the baryon
  density of the universe. There is a sensitive velocity dependence
  for all mechanisms of baryon production, depending on the magnitude
  of velocity of the bubble wall, and we examine in particular an
  inverse velocity dependence on baryon production, which is predicted
  by the charge transport mechanism of baryon production. For this
  mechanism we find both an enhancement of baryon production and the
  generation of inhomogeneities during the electro-weak phase
  transition. We calculate the magnitude of the baryon enhancement,
  which can be as large as a few orders of magnitude, depending on the
  parameters of the theory, and we calculate the size and amplitude of
  the inhomogeneities generated. We determine that the inhomogeneities
  generated in a thermally nucleated electro-weak phase transition are
  to small to survive diffusive processes and effect the
  nucleosynthesis epoch. We also examine the possibility that a phase
  transition nucleated by other means, such as by the presence of
  cosmic strings, may produce inhomogeneities that could effect
  nucleosynthesis.
\end{abstract}
\footnotetext[1]{Present address is at U. of Washington,
    aheck@ptolemy.astro.washington.edu, but beginning 9/94 the address
    is with the Cosmology/Astrophysics group at Fermilab.}

\section{Introduction}

If a phase transition occured in the early universe, would it leave
behind any remnants that we can observe today? The theory of
electro-weak baryogenesis proposes that an electro-weak (EW)
phase transition left behind the most observable remnant of all:
baryons.

The present problem with the theory of EW baryogenesis, however, is
that it has too many unknown parameters. For this reason, we are
presently unable to determine whether the theory agrees with observed
cosmological baryon density or not. For example, in making a
calculation of the baryon density, one must leave some unknown
parameters (such as those depending on the origin of $CP$ violation)
in the final answer. Factors of order unity that are due to
theoretical uncertainties are also left in the final answer.
Certainly, calculations of theory have shown that EW baryogenesis is
{\em possible}: one can simply adjust the unknown parameters (within
reason) to make the theory match observation, but whether or not the
actual values of the unknown parameters correctly predict the observed
baryon density, is still not known. Improvements must still be made on
both the theoretical calculations and on experimental observations
(i.e.  via particle accelerators) of the unknown parameters.

In this paper, we will explore both the behavior of the phase
transition and the production of baryons during the phase transition
for large regions of this parameter space. By examining a detailed
picture of phase transition dynamics, we will find that the evolution
of the phase transition is very sensitive to some of the parameters of
EW theory, such as latent heat and correlation length. Once we are
armed with a consistent picture of phase transition dynamics, we will
find that we can make two important additions to the calculation of EW
baryogenesis.

The first addition is a numerical correction to the calculation of the
average baryon density produced in the EW phase transition, that can
be as large as a few orders of magnitude. This correction, which has
been previously neglected, originates from the observation that, for
many models, baryon production is sensitively dependent on the
velocity of the bubble walls in the phase transition.  By including a
more complete picture of phase transition dynamics, which includes
reheating of the plasma from released latent heat and the deceleration
of the bubble walls due to this reheating, we will find that the
baryon production can be enhanced by a large factor for a large range
of parameters of the EW theory. This enhancement is independent of
parameters such as the strength of $CP$ violation, so the
observational constraints on these factors can be relaxed.

The second addition addresses the possibility of {\em adding another
  observational constraint from cosmology}: the primordial abundances
of the elements. We will find that, if baryon production is velocity
dependent, as many theories predict, then large amplitude
inhomogeneities are produced during the phase transition for a large
range of EW parameters.  Inhomogeneities in the early universe are
important because measurable primordial abundances of the elements are
sensitive to certain sizes and amplitudes of inhomogeneities present
at the time of Big Bang nucleosynthesis.  That is to say, the EW phase
transition may have left its mark not only on the number of baryons it
produced, but also on the kinds of elements that were formed during
Big Bang nucleosynthesis.  We will see that baryon production and the
amplitude of the inhomogeneities are sensitive functions of the (as
yet) unknown parameters of EW theory.  In this light, another way of
looking at this model is to ask the question: Can certain regions of
EW parameter space be ruled out, not only by careful measurements of
the baryon density of the universe, but also by constraints on
inhomogeneities allowed by Big Bang nucleosynthesis calculations and
observations? We will find that the inhomogeneities produced in a
thermally nucleated EW phase transition are too small to survive
important diffusive processes, and so the inhomogeneities do not play
a significant role in Big Bang nucleosynthesis. However, we also
examine the possibility of non-thermal nucleation (seeding) of the
phase transition, which allows larger size inhomogeneities that can
affect Big Bang nucleosynthesis.

On our way to calculating a more complete picture of the phase
transition dynamics and ultimately coupling this to baryon production,
we will address several important issues. For example, we must justify
our assumption that baryon production is sensitively dependent on the
velocity of the bubble walls. Also, since the velocity of the bubble walls
is crucial not only for the calculation of baryon production but also
for the calculation of the phase transition dynamics, we will
investigate in detail a method for calculating the velocity of the
bubble walls which uses a damping coefficient $\eta$ as a free
parameter \cite{Ignatius94a}. In this calculation of wall velocity, we have
included several important effects previously neglected by many
authors.

We should mention here that, so far, we have referred to ``the'' EW
theory, when in fact there are many different EW theories. For example
there are EW theories which include several scalar particles which are
involved in the phase transition, and there are also EW theories which
are part of of bigger super-symmetric theories (see \cite{Cohen93a}
for a review). However, we will find that the calculations in this
paper are not sensitive to the particular type of theory chosen. The
only assumption we make is that the baryon production is sensitive to
the velocity of the bubble wall, and we will see that this is true
independent of the particular EW theory used. Since at times we will
need to use specific examples in order to get numerical results, we
will use what is called the `minimal' standard model EW theory.
Therefore from now on in this paper when we refer to the EW theory, we
mean the minimal standard model of EW theory, keeping in mind that
our conclusions will still apply for other theories.

The structure of the paper is as follows: First of all, the reader who
is only interested in the final results of the evolution of the phase
transition, the resultant baryon production, and the effects on
nucleosynthesis can refer directly to sections~5, 6 and 7. In
section~5, we discuss three different possibilities in which the phase
transition can proceed. In each case we produce formulas for the
evolution of the phase transition, most importantly the evolution of
the bubble wall velocity, for this is what determines baryon
production. In section~6 we present the numerical results for the
(non-trivial) cases of evolution of the phase transition. We calculate
the total baryon density produced (relative to the constant bubble
wall velocity calculation), and the size and amplitude of the
inhomogeneities produced for the non-trivial cases, each depending on
the parameters of EW theory. In section~7 we discuss how the
inhomogeneities generated in the phase transition can effect Big Bang
nucleosynthesis. We review the effect of diffusive processes on
inhomogeneities in the early universe, and we show that the
inhomogeneities generated in a phase transition initiated by thermal
nucleation, are dissipated by baryon diffusion before they can have
any effect on nucleosynthesis. We also discuss the possibility of
non-thermal nucleation producing larger size inhomogeneities.

In sections~2, 3, and 4 we discuss the important detailed calcualtions
that lead up to our results. For example in section~2,  we briefly
review the importance of the EW phase transition in the theory of EW
baryogenesis, and then we investigate in detail how baryon production
depends on the velocity of the bubble walls nucleated in the phase
transition. In section~3 we apply the EW theory to phase transition
dynamics in order to express all of the important quantities of the
phase transition, such as bubble nucleation rate and fraction of space
converted to the low temperature phase, in terms of the parameters of
the EW theory. We also produce a formula for determining the average
baryon density produced during the phase transition. In section~4 we
discuss the propagation of the bubble wall in detail. We introduce a
bubble wall damping coefficient $\eta$, and we include hydrodynamics
to obtain a formula for the bubble wall velocity which is an explicit
function of the parameters of EW theory, $\eta$, and the temperature
of the plasma on both sides of the bubble wall. We then solve for the
bubble wall velocity by using conservation of energy-momentum and
reasonable boundary conditions for temperature and bulk fluid flow,
and obtain wall velocities that are functions of EW parameters and
$\eta$ only.

Finally the last section summarizes the main conclusions of this
paper.

\section{Electro-weak Baryogenesis}

The quantum field theoretical aspects of EW baryogenesis are well
developed, so we refer the interested reader to a recent review of EW
baryogenesis by Cohen, Kaplan, and Nelson \cite{Cohen93a}. The
important point that we wish to use from the study of EW Baryogenesis
is that the third of Sakharov's conditions for baryogenesis, namely
that baryogenesis requires an out-of-equilibrium environment, is met
via the EW phase transition \cite{Kuzmin85a}.

Let us describe the EW phase transition. First of all, whether or not the
EW phase transition is first or second order (or if there is any
transition at all) is still an unresolved matter \cite{Cohen93a},
though the present bias is towards a first order transition. In this
paper, we will take the tack of {\em assuming a first order phase
  transition} because this is the type that has the greatest chance of
leaving observable remnants and is therefore the most verifiable. A
first order phase transition proceeds in the following way.  When the
universe is at a temperature above the critical temperature $\Tc$, the
plasma is in its high temperature unbroken (``u'') phase. As the
universe expands, it cools down below the critical temperature and
bubbles of the low temperature broken (``b'') phase begin to nucleate.
Once a bubble has nucleated, it will begin to grow and its bubble wall
velocity will quickly reach some velocity $v_{0}$, which depends on
the temperature of the plasma and the internal bubble wall dynamics.
As the bubbles grow and convert the plasma to the low temperature
phase, they will also release a certain amount of latent heat $L$,
which is an (as yet) unknown parameter of EW theory, and this will tend to heat
the plasma towards $\Tc$. This reheating will in turn decrease the
bubble wall velocity, which goes to zero as $T \rightarrow \Tc$. If $L
\ll \rho(\Tc)-\rho(\Tn)$, where $\Tn$ is the temperature at which the
bubbles nucleate and $\rho$ is the energy density, then the bubble
walls will only slightly slow down.  If $L \sim \rho(\Tc)-\rho(\Tn)$,
then we will show that the bubble wall will slow down drastically
($v_{0}\ll 1$) until the expansion of the universe itself can absorb
the latent heat \cite{Kajantie86a}.  Of course the picture is more
complicated than this because the reheating is not homogeneous, but
qualitatively this description is accurate.

The bubble walls meet Sakharov's third and final requirement for
baryogenesis: the plasma in and immediately around the bubble wall is
out of equilibrium. Therefore the bubble wall plays an important role
in baryon production. However, baryon production need not occur in the
bubble wall itself. For example, transport processes can cause the
region in front of the bubble to be out of equilibrium, thus allowing
baryogenesis to occur in front of the bubble wall
\cite{Nelson92a,Joyce94a}. There are other mechanisms, such as
spontaneous baryogenesis, which produce baryon number inside the
bubble wall \cite{Cohen91a,Dine94a}. Exactly which mechanism
dominates, and where most of the baryons are created, depends upon
parameters such as the bubble wall thickness and bubble wall velocity
\cite{Cohen93a,Joyce94a}.

The dominant baryon production mechanism will in turn more precisely
determine how baryon production depends on bubble wall velocity. We
will see that the velocity dependence is important for determining the
effects of phase transition dynamics on the total baryon density and
on the production of inhomogeneities. So let us now examine the
velocity dependence of baryon production more closely.

\subsection{Bubble wall velocity and baryogenesis}

The crucial assumption of this paper is that baryon production rate is
a sensitive function of bubble wall velocity for at least some range
of velocities.

What is our justification for this assumption? How can we even justify
that baryon production has {\em any} dependence on the bubble wall
velocity? Strictly speaking, there is no question that baryon
production in the EW phase transition depends on the bubble wall
velocity $v_{0}$.  How sensitive a function it is of velocity,
however, depends on both the (dominant) mechanism of baryon production,
and the magnitude $v_{0}$.  For example, in the limit that
$v_{0}\rightarrow 0$ (and keeping the bubble wall thickness constant)
the plasma in and near the bubble wall remains in thermal equilibrium,
thus no baryon number is produced, independent of the mechanism of
production. As the $v_{0}$ increases from zero, however, we will see
that the velocity dependence of baryon production depends on which
mechanism of baryon production is dominant.

In order to determine which mechanism of baryon production is
dominant, one must compare three different time scales \cite{Cohen93a}: the
thermal time scale $\tau_{T}\sim$ few$/T$, the baryon violation
time scale $\tau_{B}\sim 1/\alpha^{4}T$, and the time scale of the wall
$\tau_{w}\sim \delta_{w}/v_{0}$, where $\delta_{w}$ is the bubble wall
thickness. Nonetheless, we will find that all baryon production
mechanisms have some common velocity dependent characteristics at
small and large velocities. Before we discuss these general
characteristics, let us first take a close look at a specific
mechanism in order to see how the bubble wall velocity dependence of
baryon production is determined in detail.

\subsubsection{The charge transport mechanism}
In order to quantitatively calculate the effects of phase transition
dynamics on baryon production, we will choose a model a baryon
production based on the charge transport mechanism, which predicts that
the baryon production is inversely proportional to the bubble wall
velocity \cite{Nelson92a,Joyce94a}. That is to say, our model for the
calculation of baryon density $n_{\rm B}(t)$ produced at some time $t$
during the phase transition consists of one equation:
\begin{equation}
n_{\rm B}(t) = \frac{{\cal A}}{v_{0}(t)}
\label{nB}
\end{equation}
where ${\cal A}$ is some constant which depends on the specific
parameters of the theory of baryon production. For example, ${\cal A}$
is proportional to the amount of $CP$ violation in the theory. Because
we are grouping all of our ignorance about these parameters in ${\cal
  A}$, we will not be able to calculate absolute numbers for total
baryon density, but only numbers relative to some particular theory of
baryon production represented by ${\cal A}$. In this way, we can
isolate the effects of phase transition dynamics on baryon production,
since previous authors neglected to take into account the changing
velocity of the bubble walls during the phase transition.

We choose to use the charge transport mechanism as a model because it
is a well developed, plausible theory which has an explicit prediction
of bubble wall velocity dependence \cite{Nelson92a,Joyce94a}.  Let us
examine in more detail the charge transport mechanism in order
determine how the velocity dependence of equation~(\ref{nB}) comes
about. Indeed, one might suspect that this velocity dependence must
break down for small velocities, since for $v_{0}\rightarrow 0$, this
formula predicts infinite baryon density. We will find, as Cohen,
Kaplan and Nelson \cite{Nelson92a} have, that eq.~(\ref{nB}) is valid
only for velocities above some cut-off velocity $v_{\rm co}$, which
we will calculate. In this paper, we will also go one step further and
determine the velocity dependence of baryon production as the bubble
wall velocity $v_{0}\rightarrow 0$.

The charge transport mechanism can be described in the following way.
As the bubble wall propagates through the plasma, some kind of charge
$Y$ (e.g.  lepton charge or hypercharge) is reflected off the bubble
wall in a $CP$ violating way back into the high $T$ phase.  This
excess charge is then converted into baryon number via the previously
mentioned baryon number violating processes continuously occuring in
the high temperature phase. At some point, the wall will again catch
up to the charge, which is slowly diffusing through the plasma, and as
the wall sweeps through the charge at this point, the baryon violating
processes will turn off (because they are exponentially suppressed in the
low $T$ phase), leaving behind a baryon remnant.

Nelson, Kaplan and Cohen \cite{Nelson92a} have shown that the rate per
volume at which the charge $Y$ is turned into baryon number, or what
we call the baryon number density creation rate $\dot{n}_{B}|_{\rm
  cr}$, at some point $z$ in space is
\begin{equation}
\left.\dot{n}_{B}\right|_{\rm cr}= \frac{{\cal
N}\Gamma_{B}}{T^{3}}n_{Y}(z-v_{0}t),
  \label{barycreat}
\end{equation}
where ${\cal N}$ is a number of order unity which depends on the type
of charge $Y$ and on the number of scalar Higgs in the theory,
$\Gamma_{B}$ is the baryon violation rate per volume, and $n_{Y}(z-v_{0}t)$
is the (steady state) charge $Y$ number density for a wall moving with
velocity $v_{0}$, for a point $z-v_{0}t$ away from the wall.

However, since baryon violation processes are continuously occuring,
any baryon number that is created will also tend to equilibrate to its
equilibrium value, which is zero. This behavior can be describe by a
`baryon number annihilation rate' $\dot{n}_{B}|_{\rm ann}$ which can be
approximated by using a simple form of the Boltzmann equation
\begin{equation}
\left.\dot{n}_{B}\right|_{\rm ann} \approx -\frac{n_{B}}{\tau_{B}},
  \label{baryannh}
\end{equation}
where, as stated before, $\tau_{B}$ is the baryon number violation
time scale.

Putting these two rates together and making the rough approximation
that $\tau_{B}^{-1}\approx \Gamma_{B}/T^{3}$, one obtains a formula for the
total change in baryon number density
\begin{eqnarray}
  \dot{n}_{B} \approx \frac{\Gamma_{B}}{T^{3}}\left({\cal
    N}n_{Y}(z-v_{0}t) - n_{B}\right) \nonumber \\
\approx \frac{1}{\tau_{B}}\left({\cal
    N}n_{Y}(z-v_{0}t) - n_{B}\right)
  \label{barydot}
\end{eqnarray}

The total baryon density at $z$ can be found by integrating
(\ref{barydot}) over time. We will examine two limits of this formula
in order to determine its behavior. The first limit is when $n_{Y}\gg
n_{B}$ and the second limit is when $n_{Y}\ll n_{B}$. We will see that
these two limits are equivalent to the large and small velocity limit
of the bubble wall.

The first limit $n_{Y}\gg n_{B}$ occurs when the average time
$\tau_{\rm ave}$ the charge $Y$ spends out in front of the wall after
being reflected (and before being caught again by the wall) is much
smaller than $\tau_{B}$. In this case, the charge $Y$ has not had very
much time to be converted into baryon number, and so $n_{B}$ is small
compared to $n_{Y}$. In this limit, one can neglect the $n_{B}$ term
in (\ref{barydot}) and integrate over time. But first, in order to do
this, one must calculate $n_{Y}(z-v_{0}t)$. Since the charges
thermalize in a few mean free paths, we can find an approximation to
$n_{Y}$ by using the diffusion equation. Since the wall is a constant
source of reflected charge flux $J_{Y} =J_{0}(v_{0})\delta(z-v_{0}t)$,
where we have explicitly noted that $J_{0}$ is a function of bubble
wall velocity, the equation of diffusion of charge $Y$ is
\begin{equation}
\dot{n}_{Y} +Dn_{Y}^{\prime \prime} = J_{0}(v_{0})\delta^{\prime}(z-v_{0}t)
  \label{diffusion}
\end{equation}
where $D$ is the diffusion coefficient. The solution to
equation~(\ref{diffusion}) is
\begin{equation}
n_{Y}(z) = \frac{J_{0}(v_{0})}{\bar{v}}e^{-v_{0}z/D}
  \label{nY}
\end{equation}
where we have used the boundary condition $n_{Y}(0)=J_{0}/\bar{v}$,
and $\bar{v}$ is the average velocity with which the reflected
particles leave the wall.

The integrated baryon density in the case $n_{Y}\gg n_{B}$, or (as
described above) equivalently $\tau_{B}\gg \tau_{\rm ave}$, is then
\begin{equation}
  n_{B}= \frac{{\cal
      N}\Gamma_{B}}{T^{3}}\int^{z/v_{0}}_{-\infty}dtn_{Y}(z-v_{0}t)=
  \frac{{\cal
      N}\Gamma_{B}}{T^{3}}\frac{D}{\bar{v}}\frac{J_{0}(v_{0})}{v_{0}^{2}}.
\label{nbint}
\end{equation}

The velocity dependence of $J_{0}(v_{0})$ is complicated(see
\cite{Nelson92a} for an explicit expression), but to first order in
$v_{0}$, $J_{0}\propto v_{0}$.  Therefore, in the limit $\tau_{B}\gg
\tau_{\rm ave}$, we can see that $n_{B}\propto v_{0}^{-1}$, just as in
eq.~(\ref{nB}).

In the second limit $n_{Y}\ll n_{B}$, however, the velocity dependence
is very different. This limit occurs when $\tau_{ave}\gg \tau_{B}$, giving the
reflected charge $Y$ plenty of time to convert to
baryon number such that $n_{Y}$ becomes small compared to $n_{B}$. In
this case, one can solve for the total baryon density by assuming that
all of the charge $Y$ has converted into baryon number, and so we
neglect the $n_{Y}$ term when integrating eq.~(\ref{barydot}). In
this limit the formula for the total baryon density is simply
$\dot{n}_{B}=-n_{B}/\tau_{B}$. Of course the exact solution of this
equation will be complicated because the charges are not distributed
evenly out from the wall, but the general behavior of the solution is
of the form
\begin{equation}
n_{B}\approx n_{B0}e^{-\tau_{\rm ave}/\tau_{B}},
  \label{nBdecay}
\end{equation}
where $n_{B0}$ is an average the baryon number density, averaged over
the inhomogeneous distribution of the initial $Y$ charges that have
been converted into baryon number. One would expect $n_{B0}$ to be
roughly the same as the baryon density calculated in the previous
limit, namely eq.~(\ref{nbint}).  Recall that, $\tau_{\rm ave}$ is the
average amount of time the charges stay in front of the wall after
reflection. This average time can be estimated by observing from
eq.~(\ref{nY}) that the distribution of $Y$ charges diffuses out an
average distance $D/v_{0}$ from the wall after reflection. Therefore
it will take the wall an average time of
\begin{equation}
\tau_{\rm ave}\approx \frac{D}{v_{0}^{2}}
  \label{tauave}
\end{equation}
before the wall catches up again with the reflected charges.

We can approximate the velocity dependence of $n_{B}$ in the
$\tau_{ave}\gg \tau_{B}$ limit by combining eqs.~(\ref{nBdecay}) and
(\ref{tauave}) and by using the approximation of setting $n_{B0}$
equal to the value of $n_{B}$ in equation~(\ref{nbint}) to obtain
\begin{eqnarray}
  n_{B}&\approx& \frac{{\cal
N}\Gamma_{B}}{T^{3}}\frac{D}{\bar{v}}\frac{J_{0}(v_{0})}{v_{0}^{2}}e^{-D/(v_{0}^{2}\tau_{B})} \nonumber \\
&\propto& \frac{e^{-D/(v_{0}^{2}\tau_{B})}}{v_{0}},
  \label{nbsmallv}
\end{eqnarray}
where we have used $J_{0}\propto v_{0}$. Notice that $n_{B}\rightarrow
0$ as $v_{0}\rightarrow 0$, as it should.

Now that we have obtained the velocity dependence of $n_{b}$ in two
different limits, for what range of velocities are these two limits
valid? First of all, Equation~(\ref{tauave}) explicitly shows that it
is completely equivalent to present the limits $\tau_{ave}\ll
\tau_{B}$ and $\tau_{ave}\gg \tau_{B}$ as lower and upper bounds on
the bubble wall velocity. For example, one can use (\ref{tauave}) to
express the limit $\tau_{ave}\ll \tau_{B}$ as $v_{0}^{2}\gg
D/\tau_{B}$. Thus, we can place a lower cut-off limit $v_{\rm co}$ on the
bubble wall velocity for which the approximation $n_{B}\propto
v_{0}^{-1}$ is valid. Recalling that $\tau_{B}\approx
1/(\alpha_{W}^{4}T)$, we obtain
\begin{equation}
v_{\rm co}^{2}\approx \alpha_{W}^{4}TD.
  \label{vminD}
\end{equation}
For velocities less than this cut-off velocity $v_{\rm co}$,
eq.~(\ref{nbsmallv}) is the valid form for $n_{B}$. Since the
diffusion coefficient is typically equal to a few mean free path
lengths $\lambda$ (and $\lambda \approx$ few$\times
T^{-1}$) \footnote{The mean free path is smaller than the usual estimate
  of $\lambda\sim (\alpha^{2} T)^{-1}$, where $\alpha$ is the relevant
  coupling, because at these high temperatures there are many
  different types of particles to scatter off of (gluons, quarks,
  etc.), and this significantly reduces $\lambda$ \cite{Cohen93a}.}, we
estimate the cut-off velocity to be
\begin{equation}
v_{\rm co} \sim 10^{-3},
  \label{vminEW}
\end{equation}
but because of the theoretical uncertainties of quantities such as
$\tau_{B}$ and $D$, the cut-off velocity could range anywhere from
$10^{-4}<v_{\rm min}<10^{-2}$. The estimate (\ref{vminEW}) will be
used as a reference value throughout the paper. For example, we will
see that this limit is important for establishing the maximum
allowable over-densities produced during the phase transition.  Later
in this paper, we will find out that {\em phase transition dynamics}
will also place lower limits on bubble wall propagation velocities,
and these limits will be of the same order as (\ref{vminEW}).

\subsubsection{The general velocity dependence of baryon production}

Besides the specific example of the charge transport mechanism, one
can make some remarks on the general characteristics of the velocity
dependence of {\em any} mechanism of baryon production. For example,
in the limit of very slow bubble wall velocities, the Boltzmann
equation for the annihilation of baryon number
$\dot{n}_{B}=-n_{B}/\tau_{B}$ will play a major role in any baryon
production mechanism. Just as we saw for the charge transport
mechanism, this leads to a decrease in the baryon number of the form
\begin{equation}
n_{B}\approx n_{B0} e^{-\tau_{\rm ave}/\tau_{B}}.
  \label{ngensmallv1}
\end{equation}
where $\tau_{\rm ave}$ is the average time  the baryon charges are
exposed to the equilibrating (towards zero) baryon violation
processes. For every mechanism,  $\tau_{\rm
  ave}$ will be some monotonically decreasing function of $v_{0}$. If
we assume that $\tau_{\rm ave}= D_{0}/v_{0}^{n}$, where $D_{0}$ is a
characteristic length scale of the particular mechanism, and the
exponent $n$ is also dependent on the particular mechanism, we find a
general behavior in the small velocity limit
\begin{equation}
n_{B}\propto  e^{-D_{0}/(v_{0}^{n}\tau_{B})},
  \label{ngensmallv2}
\end{equation}
which goes to zero as $v_{0}\rightarrow 0$. Note that, as in the
charge transport mechanism, there could also be a velocity dependent
coefficient (of some power of $v_{0}$) multiplied onto this
exponential dependence, but the exponential is the dominating behavior
of $n_{B}$ at small velocities.

In the large velocity limit $v_{0}\rightarrow 1$, baryon production
must also go to zero for every mechanism. This is because, for a given
bubble wall thickness $\delta_{w}$, the time
$\tau_{w}=\delta_{w}/(\gamma v_{0})$ that a given point in the plasma
experiences out of equilibrium conditions will go to zero as
$v_{0}\rightarrow 1$ (due to Lorentz contraction of the wall), and
when $\tau_{w}\ll \tau_{B}$, the velocity will be so fast that
the baryon violation processes will not have time to create any baryons.

Putting these two limits together, we conclude that for {\em every}
mechanism of baryon production, there is a velocity range for which
$n_{B}$ monotonically increases with $v_{0}$, and and there also a
velocity range for which $n_{B}$ monotonically decreases with $v_{0}$.
And for the case of small velocity, equation~(\ref{ngensmallv2})
indicates that $n_{B}$ is an exponentially sensitive function of
$v_{0}$. Figure~1 schematically presents the general
velocity behavior for all mechanisms.

In summary, all models of bubble wall baryogenesis are dependent on
the velocity of the bubble wall, and in at least some range of
velocities they are {\em sensitively} dependent on bubble wall
velocity. We will find that, for some regions of parameter space,
$v_{0}$ varies by a couple orders of magnitude during the phase
transition. Therefore it is very plausible that $v_{0}$ will be in a
range for which baryon production is sensitively velocity dependent,
and phase transition dynamics will play an important role in baryon
production.  For this paper, we will concentrate on a model with a
$v_{0}^{-1}$ dependence because it is a generic feature of a wide
class of models that work via the charge transport mechanism. However,
we will make comments on a linear $v_{0}$ dependence also. One should
keep in mind that since the phase transition dynamics are separate
from baryon production, it is a simple matter to apply {\em any}
velocity dependence of baryon production to the phase transition and
calculate the average density and type of inhomogeneities produced.

\section{Phase Transition Dynamics}

In this section, we review the theory that describes the initial onset
and subsequent evolution of the EW phase transition. To begin
properly, we must first discuss the dynamics of the order parameter
which governs the phase transition.

\subsection{The Higgs field potential}

The scalar Higgs field $\phi$ is the order parameter of the phase
transition.  The behavior of the Higgs field in a first order phase
transition has been investigated by many authors (see for example,
\cite{Coleman73a,Linde83a,Dine92a,Anderson92a} and references in
\cite{Cohen93a}). At high temperature the vacuum expectation value of
the Higgs field is zero because this is the unique equilibrium state.
For sufficiently light Higgs mass, the minimal EW theory predicts
that, as the temperature decreases, the particles in the plasma interact in
such a way that a second stable state (with non-zero vacuum
expectation value) for the Higgs appears, though it is in a higher
energy state.  At some critical temperature $\Tc$, this second stable
state has the same energy as the high $T$ state, and for $T< \Tc$, the
second stable state has a lower energy that the high $T$ state.  Because it is
energetically favorable, the field will
eventually make the transition to the second state and thus acquire a
non-zero vacuum expectation value, but first it must cross an energy
barrier between the two states. The phase transition therefore
corresponds to the passage of the vacuum expectation value of the
Higgs field from zero to a non-zero value.

In order to model the phase transition, we must first describe the
scalar field potential. We will assume the generic form
\cite{Linde83a,Dine92a,Anderson92a}
\begin{equation}
V(\phi,T)=
A(T^{2}- \To^{2})\phi^{2}-BT\phi^{3}+\frac{\lambda}{4}\phi^{4}.
\label{V}
\end{equation}

This approximation is chosen because of its simplicity, and because
it contains all of the parameters necessary to describe the
qualitative features of the potential. That is to say, this
potential can describe phase transitions ranging all the way from
strongly first order to second order.  Much work has been done to
determine the precise form of the potential (e.g.
\cite{Shaposhnikov92a,Carrington92a,Arnold93a,Dine92a}). For example,
many authors have used perturbation theory and have found logarithmic
corrections to (\ref{V}). Perturbation theory, however, breaks down
for certain regions of parameter space, namely for large coupling
$\lambda$ (or equivalently, large Higgs mass), which seems to be the
relevant part of parameter space, according to experimental lower
limits on the Higgs mass. The potential (\ref{V}) also neglects
contributions from finite mass effects of the particles of the theory
(most notably the top quark, which is the heaviest besides the Higgs)
\cite{Anderson92a}. Even with all of these corrections, the potential
must still take the general shape of two minima separated by a barrier
in order to describe a first order phase transition, so we will use
the form (\ref{V}), keeping in mind that small differences in the
numerical results will arise if one chooses a slightly different shape
for the potential.

The high temperature limit of the potential calculated from one-loop
perturbation theory does take the from of (\ref{V}), and the
parameters $A,B, \To,$ and $\lambda$ (which is weakly temperature
dependent) can be put in terms of the zero temperature masses of the
Higgs, top quark, etc.  Instead of this parameterization, we will
choose to express $A,B, \To,\lambda$ in terms of four parameters which are
more physically descriptive of the phase transition: The latent heat
$L$, the surface energy of the wall $\sigma$, the Higgs correlation length
$\xi$ (to leading order), and the critical temperature $\Tc$
\cite{Carrington93a,Ignatius94a}. This is a completely
equivalent way of parameterizing (2), and these parameters are defined
in the following way:
\begin{eqnarray}
L   & \equiv & T \left. \frac{\partial V(\phi,T)}{\partial T}
\right|_{\phi_{0},
\Tc}=\frac{8\To^{4}A^{2}B^{2}}{(\lambda A- B^{2})\lambda};\nonumber \\
\Tc & \equiv &  \frac{\To}{(1-B^2/\lambda A)^{1/2}}
\nonumber \\
\xi & \equiv & \left(\left. \frac{\partial^{2}V(\phi,T)}{\partial
  \phi^{2}}\right|_{\phi_{0},\Tc}\right)^{-1/2}=\frac{\sqrt{\lambda
  A-B^{2}}}{\sqrt{2A}\To B}\nonumber \\
\sigma & \equiv & \left.\int \left(\frac{\partial \phi}{\partial

x}\right)^{2}dx\right|_{\Tc}=\frac{2\sqrt{2}}{3}\frac{B^{3}\To^{3}A^{3/2}}{\lambda(\lambda A-B^{2})^{3/2}}
\end{eqnarray}
where $\phi_{0}=2B\To\sqrt{A}/(\sqrt{\lambda(\lambda A-B^{2})})$ is
the value of the Higgs field (at $\Tc$) at the second (low $T$ phase)
minimum, and we calculated the function $\phi(x)$ using $V(\phi,T)$ in
the equations of motion for $\phi(x)$. We will talk more about
$\phi(x)$ in later sections.

In order to get an idea of the size of these parameters, one can refer
to the perturbation theory (one-loop) calculation of the Higgs
potential. In this theory, one can calculate the four parameters
simply by knowing the mass of the $Z$ and $W^{\pm}$ bosons, the mass
of the top quark (the heaviest quark), and the mass of the Higgs
\cite{Dine92a}. The mass of the $Z$ and $W^{\pm}$ are known, $M_{Z}\simeq 91$
GeV, $M_{W^{\pm}}\simeq 80$GeV.  The mass of the top is presently
`known' (it has yet to be reliably detected) to about 20\%,
$M_{t}\approx 175$GeV. The mass of the Higgs has an experimental lower
bound of $M_{H}>60$GeV. Using these values (including $M_{H}=60$GeV),
one obtains
\begin{eqnarray}
L_{0}&\approx& 0.14\:T_{c0}^{4} \:\:\:\:\:\:\:\:\:\: T_{c0}\approx 93.7\:{\rm
GeV} \nonumber \\
\xi_{0} &\approx& \frac{12.7}{T_{c0}} \:\:\:\:\:\:\:\:\:\:\:\:\:  \sigma_{0}
\approx 0.0056\:T_{c0}^{3}
\label{EWparams}
\end{eqnarray}
where the subscript ``0'' stands for the one-loop calculation values.
We will use these values as a reference point throughout the paper,
though the whole point of this paper is to investigate all values of
parameter space. Unless, otherwise stated, these parameters in this
paper will take these one-loop values.

\subsection{Bubble nucleation}

Now that we have the Higgs potential, we can calculate the mechanics
of the phase transition. As the plasma cools down, the Higgs field,
which is initially in the high temperature unbroken phase, will
become meta-stable, and it will eventually decay from the high $T$
phase to the low $T$ phase. The decay to the low $T$ phase is done via
bubble nucleation.

A systematic theory of bubble nucleation from a meta-stable state has
been developed by Langer \cite{Langer69a}. Others have applied it to
cosmological phase transitions
\cite{Coleman77a,Linde83a}. These theories show that the
probability of crossing over to the low $T$ state per unit time per
unit volume is \cite{Linde83a}
\begin{equation}
\Gamma_{\rm nuc} \sim T^{4}e^{-S_{3}(\phi_{cl},T)/T}
\label{Gamapp}
\end{equation}
where $S_{3}(\phi,T)$ is the three-dimensional action
\begin{equation}
S_{3}(\phi,T) = \int d^{3}x[\frac{1}{2}(\nabla \phi)^{2} + V(\phi,T)]
\end{equation}
and $\phi_{cl}$ in (\ref{Gamapp}) is the solution to the equation of motion
\begin{equation}
\frac{d^{2}\phi}{dr^{2}} + \frac{2}{r}\frac{d\phi}{dr} =
\frac{\partial V(\phi,T)}{\partial \phi},
\label{phidiff}
\end{equation}
where $r$ is the radial distance from the center of the bubble, and
the field $\phi_{cl}$ must satisfy the boundary conditions
$\phi_{cl}(r=\infty)=0$ and $d\phi_{cl}/dr|_{r=0}=0$. We should note
here that eq.~(\ref{Gamapp}) is actually the high temperature limit to
the nucleation rate. This limit is accurate if $1/T$ is smaller than
other scales in the problem such as initial bubble radius and wall
thickness. Since $1/T$ is much smaller than these scales for the EW
phase transition, this is a good approximation.

Specifically, we will use the formula derived by Linde \cite{Linde83a}
\begin{equation}
\Gamma_{\rm nuc}(T) = \kappa T\left(\frac{S_{3}(\phi,T)}{2\pi
T}\right)^{3/2}e^{-S_{3}(\phi,T)/T}
\end{equation}
for nucleation rate per unit volume. The constant $\kappa$ is thought to be
of ${\cal O}(1)$, so we will set $\kappa=1$.

Unfortunately, the solution to equation~(\ref{phidiff}) has no known
analytical form, so the calculation of $S_{3}(\phi,T)$ must be done
numerically.  Dine et al \cite{Dine92a} have, however, found an
approximation to $S_{3}(\phi,T)$ that takes a simple analytical form. This
approximation assumes that $V(\phi,T)$ is of the form (\ref{V}), and
$S_{3}$ becomes a function of two parameters $\alpha$ and $T$,
\begin{equation}
\frac{S_{3}(\alpha,T)}{T} \approx
\frac{4.85(2A(T^2-\To^2))^{3/2}}{B^{2}T^{3}}\left(1+\frac{\alpha}{4}\left(1+\frac{2.4}{1-\alpha}+\frac{0.26}{(1-\alpha)^{2}}\right)\right),
\end{equation}
where
\begin{equation}
\alpha = \frac{\lambda(2A(T^2-\To^2))^{2}}{2B^{2}T^{2}}.
\end{equation}
This function is accurate to a few percent in the whole interval of
$0<\alpha <1$ (i.e. $T_{0}<T<\Tc$). We will use this approximation in all of
the following
calculations of the phase transition.

The size of the nucleated bubbles can be determined by finding the
solution to (\ref{phidiff}) with the same boundary conditions
mentioned after the formula. This has been done by Dine et al \cite{Dine92a},
and they find that a typical nucleation radius $R_{\rm nuc} \sim
10$--$100(1/T)$.

\subsection{Phase Transition Dynamics: the middle game.}

Once the temperature of the universe drops below the critical
temperature $\Tc$, bubbles of the new phase will begin to nucleate
randomly in space, and, as we will see in the next section, each bubble
will immediately start to grow and fill up space. In order to
calculate the volume of space occupied by the new phase, we will
follow the prescription of Guth and Weinberg \cite{Guth81a}, where
they find that at some time $t$, the fraction of space $F(t)$ left in
the high temperature unbroken $u$ phase is
\begin{eqnarray}
F(t)= \exp\left\{-\frac{4\pi}{3}\int^{t}_{t_{c}}dt_{1}\Gamma_{\rm
nuc}(t_{1})\left[\int^{t}_{t_{1}}dt_{2}v(t_{2})\right]^{3}\right\}
\label{F}
\end{eqnarray}
where $v(t_{2})$ is the velocity of the bubble wall at time $t_{2}$.
Note that we have neglected the the initial radius of the bubble and
the acceleration up to terminal velocity, because both of these
effects are very small. We have also excluded the effects of the
expansion of the universe \cite{Linde83a} because for all cases in
this paper, we will see that the duration of the phase transition is
much less than the Hubble time. This formula for the fraction of space
occupied does not take into account the dynamics of the colliding
bubble walls, nor does it adequately describe the final moments of the
transition (e.g.  $F$ never reaches zero). One should therefore keep in
mind that equation~(\ref{F}) is only a convenient approximation (see
\cite{Ignatius93a} for a discussion).

In order to calculate $F$, we need to know $v(t)$. However, in order
to find $v(t)$, we need to keep several issues in mind. First, we will
see in the next section that the bubble wall velocity is temperature
dependent. More specifically, as the bubble wall is propagating, it is
releasing latent heat energy, which will heat up the plasma. One wall
itself will just reach a terminal velocity, but when two walls
approach each other, they will change each other's temperature
profile, and their velocities will change. In fact, since the walls
always heat the plasma, we will see that the wall velocities will
always decrease when the bubbles influence each other, and if the
latent heat $L$ is large enough, the bubble walls will slow down to
(almost) zero velocity.  Commonly, it is thought that since $L$ is so
small for the EW phase transition, this reheating will have a
negligible affect on the dynamics. But the important number is the
{\em ratio} $L/(\rho(\Tc)-\rho(T))$ where $\rho$ is the energy density
of the plasma. Since the super-cooling is also small for the EW
transition, this ratio can be greater than unity, and the velocities can
decrease by as much as a few orders of magnitude.

Therefore, since the velocity depends on the temperature, and the
temperature depends on how much latent heat is released, which in turn
depends on $F$, we can see that the calculation of $F(t)$ and $v(t)$
requires a full analysis of the phase transition. If one also takes into
account the fact that the bubble walls are heating the plasma
inhomogeneously, it is easy to see that an analysis of the phase
transition can become even more complicated.

In order to simplify matters, we will consider three different
scenarios of the phase transition. Each scenario is distinguished by
the initial velocity of the bubble wall (that is, before being
influenced by others). Before we investigate these three scenarios,
however, we must first investigate bubble wall propagation in order to
determine how the velocity of the bubble wall depends on temperature.
This investigation will also give us a guide to which scenario of the
phase transition should apply, for given values of the free
parameters of EW theory. We will do this in the next section. Before
going on to this, let us finish this section by discussing
specifically how we will calculate the baryon density produced during
the phase transition.

\subsection{Calculating the baryon density}

One of the goals of this paper is to calculate the effect of the
varying bubble wall velocity on the average baryon density of the
universe. Actually, we will not calculate the total baryon density,
but rather the {\em ratio} of the total baryon density produced using
a formula which accounts for variation in the bubble wall velocity,
divided by the total baryon density produced assuming the bubble wall
velocity is constant (which has been the standard assumption). The
formula for this ratio is derived in the following way.

Let us define $dB$ as the number of baryons produced in some volume
element $dV$, where $V= (1-F(t))V_{0}$ is the volume of space that has
been converted to the low temperature $b$ phase in some fiducial
volume $V_{0}$\footnote{Since the time and distance scales for baryon
  production are much smaller than the macroscopic time and distance
  scales that the baryon density varies, the use of differentials is a
  good approximation.}. Since $n_{\rm B}=dB/dV$, we can use
equation~(\ref{nB}) to obtain
\begin{equation}
\dot{B}= -\frac{{\cal A}}{v(t)}\dot{F}V_{0},
\label{B}
\end{equation}
where we have assumed that the expansion of the universe is slow
compared to the phase transition. We can then define a ``baryon
enhancement ratio'' $\chi$ as the total number of baryons produced in
$V_{0}$, accounting for the varying bubble wall velocity, divided by
the number of baryons produced assuming constant wall velocity:
\begin{equation}
\chi \equiv \frac{B(v(t))}{B(v=v_{0})}= -\int
\frac{v_{0}}{v(t)}\dot{F}(t)dt,
\label{chi}
\end{equation}
where $v_{0}$ is the velocity of the bubble wall before it has been
influenced by the presence of other bubbles. $B(v=v_{0})$ is the
number of baryons produced assuming the bubble wall velocity is
constant at some $v_{0}$.

The explicit form of $\dot{F}(t)$ is
\begin{eqnarray}
\dot{F}(t)&=& -4\pi v(t)
\left(\int_{t_{c}}^{t}dt_{1}\Gamma_{\rm
nuc}(t_{1})\left[\int^{t}_{t_{1}}dt_{2}v(t_{2})\right]^{2}\right)
\nonumber \\
& &\times \exp\left\{-\frac{4\pi}{3}\int^{t}_{t_{c}}dt_{1}\Gamma_{\rm
nuc}(t_{1})\left[\int^{t}_{t_{1}}dt_{2}v(t_{2})\right]^{3}\right\}.
\label{Fdot}
\end{eqnarray}

Once we have a consistent calculation of the evolution of the phase
transition, we can then apply eq.~(\ref{chi}) to find how the phase
transition dynamics effect the final average baryon density of the
universe. The numerical results of eq.~(\ref{chi}) are presented in
section~6.

\section{Bubble wall Growth}

As seen from the last section, the determination of the bubble wall
velocity is necessary for calculating both the phase transition
dynamics and baryon production. Specifically, we must calculate the
bubble wall velocity both before and during bubble wall collisions.
These calculations are compounded by the following complication.

According to the laws of conservation of energy and momentum, there
are two different ways a bubble wall can propagate: either as a
deflagration or detonation. Simply put, for a growing bubble of low
temperature phase, nucleated in a plasma at rest, a deflagration is a
propagating phase boundary preceded by a shock, and a detonation is a
propagating phase boundary with no shock preceding it, but a
rarefaction wave following it. For a complete discussion of these two
cases see references
\cite{Landau82,Courant85a,Kurki-Suonio85a,Laine94a}. For a schematic
presentation of the two cases, see figure~2.

Whether the bubble actually propagates as a detonation or a
deflagration, however, is unambiguously determined by the internal
dynamics of the bubble wall itself.  Unfortunately, our knowledge of
the EW phase transition is not detailed enough to reliably determine
the internal dynamics of the bubble wall
\cite{Liu92a,Dine92a,Arnold93b}, so we are still unsure whether it
propagates as a detonation or deflagration. One can, however {\em
  parameterize} the internal dynamics of the wall (e.g. see
\cite{Ignatius94a}).  This is exactly what we will do in this section.
We will employ a single parameter, in the form of a damping
coefficient $\eta$, to describe the internal dynamics of the bubble
wall.

This parameterization will allow us to determine which regions of EW
(and damping coefficient) parameter space produce deflagrations and
which produce detonations. It is important to distinguish between
these two types of bubble wall propagation because, as we will see in
the next section, they produce large quantitative and qualitative
differences in the phase transition dynamics. We will concentrate on
bubble walls propagating as deflagrations because we will find that a
large region of parameter space (which easily includes the `expected EW
values' of the parameters according to one loop calculations),
predicts that the walls will propagate as deflagrations. It is also
the case that deflagrations produce the most interesting phase
transition dynamics.

The parameterization of the internal dynamics of the bubble wall will
also allow us to calculate the velocity of the bubble wall as a
function of temperature, and this will be important when dealing with
bubble collisions. First, though, let us begin with a general
explanation of bubble wall growth.

\subsection{Growth of the Bubbles}

Once a bubble has nucleated, it will start to grow because the
pressure inside the bubble in the low $T$ phase is greater than in the
high $T$ phase, and this difference is great enough to overcome the
surface tension in the bubble walls. At first, the bubble walls will
accelerate. If the bubble wall were `in a vacuum', then we could use
(\ref{phidiff}) to estimate that the wall velocity would approach the
speed of light in a time scale of $\tau \sim
\phi_{0}^{2}/(\delta_{w}V(\phi_{0},T))$, where $\delta_{w}$ is the
wall thickness. Typically for the EW transition this translates into
$\tau \sim {\cal O}(10^{2}-10^{4})(1/T)$, which shows that the bubble
walls reach terminal velocity within a few hundred bubble wall widths,
easily by the time it is macroscopic (and recall from the Friedmann
equation that the Hubble length $H^{-1}$ at 100GeV is $H^{-1} \sim
m_{\rm pl}/T^{2}\sim 10^{17}(1/T)$, where $m_{\rm pl}$ is the plank
mass).  The additional coupling of the Higgs field to the thermal
plasma does not change the above conclusion that once a bubble is
nucleated, its wall will quickly reach some terminal velocity $v_{0}$
(in fact, if $\phi$ is strongly damped, the wall will reach terminal
velocity even faster). We will therefore ignore this initial
acceleration stage of the bubble.

The important question is, what is the (terminal) velocity of the
bubble wall? Presently, the answer is not known. The difficulty lies
in the fact that the velocity of the wall is determined by in internal
dynamics of the bubble wall, namely the interaction of the wall with
the plasma.

To get a physical idea of bubble wall propagation, consider the
following picture. As the wall sweeps through a specific point in the
the plasma, $\phi$ acquires a vacuum expectation value, and the
particles that are coupled to $\phi$ (quarks, etc) will acquire a mass.
If $\phi$ changes fast enough, the particles which initially had a
thermal distribution appropriate for their mass-less state, will be out of
thermal equilibrium because they now have a mass. The particles that have
an initial energy lower than the new mass they should acquire cannot
pass into the low $T$ phase, but rather are reflected, because they do
not have enough energy to pass through the wall. The reflected
particles will thus impart momentum to the wall and slow it down, and
the faster the wall goes, the more momentum the particles will impart
on the wall. With this picture, we can see that when the bubble wall
propagates through a plasma, a frictional type of force opposes the
motion.

However, if $\phi$ changes sufficiently slowly in time, then the particles
will always be in thermal equilibrium in the wall and there will be no
reflection and no velocity dependent force \cite{Turok92a}. The
two parameters that determine how fast $\phi$ changes with time are
the velocity of the wall $v_{0}$ and the bubble wall thickness
$\delta_{w}$. If the time scale for the change in $\phi$ is much
greater than the thermal time scale, $\delta_{w}/v_{0}\gg \tau_{\rm
  th}$, then the velocity dependent damping will turn off.

The above is the standard picture for analyzing the wall velocity
\cite{Dine92a}.  More precisely, to estimate the terminal velocity, one
uses the above picture to derive a formula which equates the pressure
difference across the wall, which drives the expansion, with the
leading order velocity dependence of the frictional damping pressure
\begin{equation}
p_{b}-p_{u} = {\cal E}v_{0}.
\label{damp}
\end{equation}
where $b$ and $u$ stand for the low (broken) and high (unbroken) $T$
phases respectively and ${\cal E}$ represents the calculation of
strength of reflection of the particles off the wall \cite{Dine92a}.
As explained above for $\delta_{w}/v_{0}\gg \tau_{\rm th}$ the
particles become more and more weakly reflected, and the right hand
side goes to zero \cite{Turok92a}. In this paper, we will make two
important changes to equation~(\ref{damp}). The first change comes
from the observation, as we will see in section(4.2), that $\delta p=
p_{b}-p_{u}$ is {\em velocity dependent}. This is tied in with the
second change: according to energy momentum conservation, there {\em
  must} be a temperature difference across the wall (see
section(4.2)), contrary to what previous authors have assumed
\cite{Turok92a,Dine92a,Liu92a} (although, for example, Ignatius et al.
\cite{Ignatius94a} have correctly included this effect). These two
changes will make a big difference in the calculation of the wall
velocity. In a sense, this is another kind of damping on the bubble
wall-- a damping due to a temperature dependent pressure difference
across the wall. We will discuss this again at the end of this
section.

Dine et al. \cite{Dine92a} has made an estimate of the bubble wall
velocity in EW theory using the above picture. Liu, McLerran et al.
\cite{Liu92a} have estimated the friction coefficient (for thin walls) by
using a more analytical approach. They add corrections to the Higgs
field from the loop expansion to the equation of motion of the Higgs
field to obtain,
\begin{equation}
\frac{d^{2} \phi}{d t^{2}} - \nabla^{2}\phi + \frac{dV}{d\phi} +\int
d^{4}x'\Sigma(x,x')\phi(x') =0
\end{equation}
where $\Sigma$ is the self energy of the Higgs field. This
extra term is responsible for the frictional force on the wall
(comparing this equation with eq.~(\ref{eqmotion}) in the next section
makes this more clear).  All the authors that have estimated the wall
velocity (e.g. \cite{Linde83a,Dine92a,Turok92a,Liu92a}) reach the same
general conclusion that thick walls have a larger velocity, than the
thin walls. The estimated velocities for thin walls are typically
$v_{\rm thin} \sim 0.1$, but the estimates for thick walls range from
$v_{\rm thick}\sim 0.1-1$.

\subsection{The parameterization of the wall velocity}

In this paper, we do not claim any value for the wall velocity
$v_{0}$, rather we parameterize $v_{0}$ with a damping coefficient
$\eta$, and explore the possible values of $v_{0}$ for a given
$\eta$. This in effect sweeps all of the messy calculations of the
internal dynamics of the bubble wall into one number $\eta$. The
parameterization of $v_{0}$ is done by the following the method.

The equation $\partial_{\mu}\partial^{\mu}\phi +\partial
  V/\partial \phi=0$ is the classical equation of motion for $\phi$
with no damping. We will assume that the motion of the $\phi$ is
damped, and that this damping is proportional to $d\phi/dt$, which is
the standard frictional damping assumption \cite{Goldstein80}. This
damping will add a term to the equation of motion $\eta d\phi/dt$.
The Lorentz invariant version of this term is $\eta
u^{\mu}\partial_{\mu}\phi$ where $u^{\mu}$ is the four velocity of the
plasma. The equation of motion then becomes
\begin{equation}
\partial_{\mu}\partial^{\mu}\phi +\frac{\partial V}{\partial \phi} + \eta
u^{\mu}\partial_{\mu}\phi=0.
\label{eqmotion}
\end{equation}
This equation is also derived by Ignatius et al. \cite{Ignatius94a} by
using the stress energy tensor of $\phi$ and the plasma, and assuming
some coupling between the two (note that we use $\eta$ as a damping
coefficient instead $1/\Gamma$ used in their paper. We have chosen
this convention because it is the standard damping factor convention).

Before we go on with the calculation of the bubble wall velocity, let
us insert here that $\eta$ can be estimated by observing from
(\ref{eqmotion}) that the amount of energy per volume per second
$\dot{\epsilon}$ generated by the frictional damping is
\cite{Goldstein80} $\dot{\epsilon}\approx \eta \dot{\phi}^{2}$.
Multiplying this by the bubble wall thickness $\delta_{w}$, one
obtains an estimate of the amount of energy generated per area per
second by the frictional damping.  This must be equal to $\delta
pv_{0}$, where $\delta p$ is the pressure on the wall due to the
frictional damping, and $v_{0}$ is the velocity of the bubble wall.
Therefore,
\begin{equation}
\eta \approx \frac{\delta p \delta_{w}}{\phi_{0}^{2}v_{0}}
  \label{etaapprox}
\end{equation}
where we have assumed that $\dot{\phi}\approx
\phi_{0}v_{0}/\delta_{w}$ and $\phi_{0}$ is the value of the
Higgs field in the broken phase. By using approximations for $\delta
p$ obtained from references \cite{Turok92a,Dine92a,Carrington93a} as
a guide, we roughly estimate that $\eta \sim g_{W}^{2}\Tc \approx
0.3\Tc$. This will be used as a reference value throughout the paper,
and, unless otherwise stated, $\eta$ will assume this value.

Getting back to calculating the bubble wall velocity, because the
bubble wall is so thin ($\ll$ radius of bubble), we can go to 1+1
dimensions. If we boost to a frame moving with the wall (velocity
$v$), and we assume that all processes have stabilized so that all
time derivatives vanish, equation~(\ref{eqmotion}) becomes
\begin{equation}
\phi''(x)= \frac{\partial V(\phi,T)}{\partial \phi} +\eta v \gamma\phi'(x)
\label{eqmotion2}
\end{equation}
where $\gamma = 1/(1-v^2)^{1/2}$. Furthermore, if we multiply both
sides by $d\phi(x)/dx$ and integrate over $-\infty<x<+\infty$ (recall
$\phi(\pm \infty)=$ constant) we obtain a formula for the velocity of the wall
\begin{equation}
v\gamma = \frac{1}{\eta}\frac{\int^{\phi_{u}}_{\phi_{b}}
      \frac{\partial V(\phi,T)}{\partial \phi}d\phi}{\int
      \left(\frac{\partial \phi(x)}{\partial x}\right)^{2}dx}.
\label{vel1}
\end{equation}
The integral in the numerator of (\ref{vel1}) can be performed
provided that $T(x)$ can be expressed in terms of $\phi(x)$. This is a
natural assumption to make, because, since changes in quantities such
as temperature and velocity of the plasma are driven by the phase
transition, one would expect that these quantities would be functions
of $\phi(x)$, which is the function that describes the evolution of
the phase transition. Therefore, we will assume that
\begin{equation}
T(x)= \frac{\Tu+\Tb}{2}+
\frac{\Tu-\Tb}{2}\left(1-2\frac{\phi(x)}{\phi_{b}}\right)
\label{Tprof}
\end{equation}
where $b$ and $u$ stand for the low (broken) and high (unbroken)
temperature phase, and $\phi_{b}$ is the value of $\phi$ at the
minimum of $V(\phi,T)$ in the $b$ phase.  Note that $\phi_{b}$ is
temperature dependent. A numerical, hydro-dynamical simulation of
bubble wall propagation, performed by Ignatius et al.
\cite{Ignatius94a}, has shown that equation~(3) is a good
approximation, however, they have found that for some regions of EW
parameter space, the temperature profile is slightly shifted with
respect to the $\phi(x)$, and $T(x)$ is also slightly different in shape
than (3). A shift on the order of the Higgs correlation
length $\xi$ can drastically change the value of the numerator of the
right hand side of (\ref{vel1}). We
will talk more about this shift in section(4.4).

The denominator of the right hand side of (\ref{vel1}) is the
surface tension of the bubble wall at $T$, $\sigma(T)=
\int(\partial \phi/\partial x)^{2}dx$. In order to calculate
$\sigma(T)$, we must first find $\phi(x)$. We will simplify matters by
assuming a specific form
\begin{equation}
\phi(x)= \frac{\phi_{b}}{2}\left(1+
\tanh\left(\frac{x}{n_{w}\xi}\right)\right)
\label{phi}
\end{equation}
where $n_{w}$ is the characteristic bubble wall thickness in units of
correlation length $\xi$. In the limit that $\Tb=\Tu=\Tc$, and
$\eta =0$, (\ref{phi}) is the exact solution to
equation~(\ref{eqmotion2}) (for $\phi(\infty)=0$,
$\phi(-\infty)=\phi_{b}$) with $n_{w}=2$. In reality, $n_{w}$ is a
function of $\eta$ and $v$, and we will keep it explicitly in the
formulas, though for now, we will assume that $n_{w}\simeq 2$.

Combining equations~(\ref{vel1},3,\ref{phi}), we obtain
\begin{eqnarray}
  \frac{v(\Tb,\Tu)}{(1-v(\Tb,\Tu)^2)^{1/2}} &=&
  \frac{ n_{w}\xi}{8 \lambda \eta}\left[3B\Tu \sqrt{9B^{2}\Tb^{2}-8\lambda
    A(\Tb^{2}-\To^{2})}\right. \nonumber \\
  & & \left.+\left((9B^{2}\Tb  \Tu-8\lambda A(\Tb
  \Tu-\To^{2})\right)+4\lambda
A(\Tu^{2}-\To^{2})\raisebox{.1in}[.1in][.1in]{\em  } \right]
\label{vel2}
\end{eqnarray}
and in the limit $\Tu=\Tb$,
\begin{eqnarray}
  \frac{v(\Tb=\Tu)}{(1-v(\Tb=\Tu)^2)^{1/2}} &=&
  \frac{ n_{w}\xi}{8 \lambda
\eta}\left[3B^{2}\Tb^{2}\left(1-\frac{\sqrt{B^{2}\Tb^{2}
    +8(\lambda A-B^2)(\Tc^{2}-\Tb^{2})}}{B\Tb}\right)\right.  \nonumber \\
    & & \left.  +12(\lambda
A-B^2)(\Tc^{2}-\Tb^{2})\raisebox{.15in}[.15in][.15in]{\em  } \right]
\label{velTb}
\end{eqnarray}
which explicitly goes to zero as $\Tb \rightarrow \Tc$. There are two
important notes about equation~\ref{vel2}. The first note
is that this formula is correct only to first order in $\Tu -\Tb$.
That is, we solved for $\phi_{b}$ assuming that the temperature was
uniform ($\Tb$). The fact that the temperature varies from the
unbroken phase to the broken phase will change the velocity by an
amount of order $(\Tu -\Tb)/\Tc$, which is small, so we neglect it.

The second important note about equation~\ref{vel2} is that if $\Tu
>\Tc >\Tb$ or $\Tb > \Tc > \Tu$, the wall will still propagate
forward! (See also \cite{Gyulassy84a,Ignatius94a}.) The physical
reason for this is that since the temperature is not uniform, the
shape of the Higgs potential $V(\phi,T)$ is slightly different than in
the uniform temperature case. The shape of the potential changes in
such a way that the broken phase can still have a lower free energy
that the unbroken phase {\em even if } $\Tb >\Tc >\Tu$ (this is easier to
imagine for the $\Tu> \Tc >\Tb$). Therefore the wall will propagate
forward because it is energetically favored. This can be explicitly
shown by inserting the formulas for $T(x)$ and $\phi(x)$ into
$V(\phi,T)$. Of course for the uniform temperature case
(eq.~\ref{velTb}), the bubble wall stops at $T=\Tc$.

Equation~(\ref{vel2}) brings us one step closer to determining the
wall velocity, but this equation only gives the velocity in terms of
$\Tb$ and $\Tu$ at the bubble wall. Given some bubble that nucleates
in a plasma with an initial temperature $\Tn$, how does one determine
$\Tb$ and $\Tu$ at the bubble wall? First, let's assume there is no
influence from neighboring bubbles (we'll add this in later). As
stated before, we can then impose conservation of energy and momentum
on the bubble wall to constrain the propagation to two possible forms,
namely detonations or deflagrations. For a given damping coefficient
$\eta$, equation~(\ref{vel2}) will allow only one or the other, and
this will then determine the temperature on both sides of the wall. In
the next section, we will concentrate on deflagration solutions.

Before we start the next section, let us just note here that ideally,
equation~(\ref{eqmotion2}) should be solved numerically along with the
equations from $\partial_{\mu}T^{\mu \nu}=0$ (as in \cite{Ignatius94a}
though they do this in (1+1) dimensions), to get $v(x),T(x)$ and
$\phi(x)$.  Instead we have used analytical approximations to get an
expression for $v$, because this will make computing the rest of the
phase transition easier.

\subsection{A more complete treatment of deflagration
  wall propagation}

Previous authors have not applied spherical bubble propagation theory
to the EW phase transition. There have been studies of planar wall
propagation \cite{Gyulassy84a,Kajantie86a,Enqvist91a,Ignatius94a},
but we will see that the spherical geometry of the bubbles plays a
dominant role in determining the characteristics of the bubble wall
propagation.  Most notably, the shock fronts become extremely weak for
spherical bubbles. We will use this fact to develop a useful and
intuitive approximation that simplifies the calculation of the bubble
wall velocity. This approximation is found to be very accurate in the
case of the EW phase transition.

Before we begin discussing the spherical bubble wall calculation, we
would like to note beforehand that we have also calculated the
deflagration velocity for planar walls. In a wide range of parameter
space, the planar wall calculation is a good approximation to the
spherical bubble deflagration, and the error is only $\lesssim$
1--$20\%$.  But for larger values of latent heat $L$, the error can be
as much as a factor of 2. As we will see, the bubble wall velocity is
important for determining  whether the walls travel as detonation
or deflagrations and also for determining how the phase transition
evolves. Therefore, it is important to make an accurate calculation of
the wall velocity.

The general picture of spherical deflagration wall propagation in a
plasma is the following \cite{Kurki-Suonio85a}. The deflagration front
is preceded by a shock front, which travels at a velocity $v_{\rm
  sh}>c_{s}$, where $c_{s}=\sqrt{1/3}$ is the velocity of sound in a
relativistic plasma.  In front of the shock, we impose the boundary
condition that the plasma is not moving, and has temperature $\Tn$,
the temperature at which the bubble nucleated. Between the shock and
the deflagration front, the plasma has a temperature $\Tu(z)$ which is
greater than $\Tn$, and bulk fluid velocity $v_{\rm fl}(z)$, where $z$
is the distance from the bubble wall. It is important to stress that
for spherical bubbles, the temperature and the fluid velocity of the
plasma in front of the wall are both a function of distance from the
bubble wall. Behind the deflagration front, we impose the boundary
condition $v_{\rm fl}=0$ and $T=\Tb$. See figure~2.

The velocity of the deflagration front is determined by the
conservation of energy-momentum, the equation of motion of $\phi$, and
the above boundary conditions \cite{Gyulassy84a,Kurki-Suonio85a}. The
conservation of energy and momentum is best applied by using the well
known property of the stress-energy tensor, $\partial_{\mu} T^{\mu
  \nu}=0$. First, consider a wall discontinuity, such as a shock or
deflagration front. For such a discontinuity, the problem not only
reduces to (1+1) dimensions, but if we boost to a frame comoving with
the front, all the time derivatives vanish.  Assuming that the plasma
is a perfect gas such that $T^{\mu \nu}= (\rho
+p)u^{\mu}u^{\nu}-pg^{\mu \nu}$, where $u^{\mu}=(\gamma,\gamma v_{r})$
is the covariant fluid velocity, the conservation of energy-momentum
gives us \cite{Gyulassy84a}
\begin{eqnarray}
(\rho_{1}+p_{1})\gamma^{2}_{1}v_{1}&=&(\rho_{2}+p_{2})\gamma^{2}_{2}v_{2}
\nonumber \\
(\rho_{1}+p_{1})\gamma^{2}_{1}v^{2}_{1}
+p_{1}&=&(\rho_{2}+p_{2})\gamma^{2}_{2}v^{2}_{2}+p_{2}
\label{rho1p1}
\end{eqnarray}
where the subscripts 1 and 2 refer to either side of the wall. Solving
for the deflagration front with the above boundary conditions, one
obtains \cite{Gyulassy84a}
\begin{equation}
v_{\rm def} = \sqrt{\frac{(p_{b}-p^{\rm def}_{u})(\rho^{\rm
      def}_{u}+p_{b})}{(\rho_{b}-\rho^{\rm def}_{u})(\rho_{b}+p^{\rm
      def}_{u})}}
\label{vdef}
\end{equation}
where $p$ and $\rho$ are defined by equation~(\ref{eos}), and $p^{\rm
  def}_{u}, \rho^{\rm def}_{u}$ are the pressure and energy density in
the $u$ phase at the deflagration front, (recall $\Tu(z)$ is a function
of distance $z$ from the bubble wall).

One can now use equation~(\ref{vel2}) and set $v(\Tb,\Tu^{\rm
  def})=v_{\rm def}$, where $\Tu^{\rm def}$ is the temperature of the
plasma in the $u$ phase at the deflagration front (recall $\Tb$ is
constant), in order to eliminate one unknown ($\Tb$). The deflagration
velocity is now a function of $\Tu^{\rm def}$ only (for a given
$\eta$,$\lambda$, etc.). What is the temperature $\Tu^{\rm def}$ of
the $u$ phase at the deflagration front? It is not a free parameter,
rather it is determined by the boundary conditions at the shock front.

That is to say, one can solve (\ref{rho1p1}) for the shock
discontinuity (with the above boundary conditions) in order to obtain
the fluid velocity of the plasma just behind the shock
\begin{equation}
v_{\rm fl}^{\rm sh}=\sqrt{\frac{(\rho^{\rm sh}_{u}-\rho_{n})(p^{\rm
      sh}_{u}-p_{n})}{(\rho^{\rm sh}_{u}+p_{n})(\rho_{n}+p^{\rm
      sh}_{u})}},
\label{vflsh}
\end{equation}
where the unshocked plasma $n$ has the same equation of state as the
unbroken $u$ phase, and $\rho^{\rm sh}_{u}\equiv \rho_{u}(T^{\rm
  sh}_{u})$, where $T^{\rm sh}_{u}$ is defined as the temperature of
the plasma at the shock front.

However, one can also find the fluid velocity at the shock by starting
with the fluid velocity at the deflagration front, and then use the
fluid motion equations ($\partial_{\mu} T^{\mu \nu}=0$, see below) to
find $v_{\rm fl}$ at the shock. These two different methods for
calculating $v_{\rm fl}$ at the shock must agree. Since $v_{\rm fl}$
at the deflagration front is a function of $\Tu^{\rm def}$, one can
make the two answers agree by adjusting $\Tu^{\rm def}$. This fixes
the value of $\Tu^{\rm def}$, and therefore $v_{\rm def}$.

The formulas for the temperature and fluid velocity between the
deflagration front and the shock are obtained by solving
$\partial_{\mu} T^{\mu \nu}=0$ for the case of spherical symmetry
\cite{Kurki-Suonio85a}.  These equations are enormously simplified by assuming
similarity solutions: $\rho=\rho(\xi)$, $p=p(\xi)$, $v_{\rm fl}=v_{\rm
  fl}(\xi)$, where, for radius $r$ of the bubble and time $t$,
$r/t=\xi$. This is a reasonable assumption, since we have assumed the
walls have reached a steady state. With this simplification, the
solutions become \cite{Kurki-Suonio85a}
\begin{eqnarray}
& & \:\:\:\:\:\:\:\:\:\:\:\:\:\:\:\:\:\:\:
\frac{1}{\rho}\frac{d\rho}{d\xi}=4\frac{\xi-v_{\rm fl}}{1-\xi v_{\rm
fl}}\frac{1}{1-v_{\rm fl}^{2}}\frac{dv_{\rm fl}}{d\xi} \nonumber \\
& &(v_{\rm fl}^{2}(3-\xi^{2})-4v_{\rm fl}\xi +
3\xi^{2}-1)\frac{dv_{\rm fl}}{d\xi}=2\frac{v_{\rm fl}}{\xi}(1-v_{\rm
fl}^{2})(1-\xi v_{\rm fl}).
\label{xieq}
\end{eqnarray}

By setting the boundary conditions $T(\xi=v_{\rm def})= T^{\rm
  def}_{u}$ and $v_{\rm fl}(\xi=v_{\rm def})= v^{\rm
  def}_{\rm fl}$, one can solve (\ref{rho1p1}) to find
\begin{equation}
v_{\rm fl}^{\rm def}=\sqrt{\frac{(\rho^{\rm def}_{u}-\rho_{b})(p^{\rm
      def}_{u}-p_{b})}{(\rho^{\rm def}_{u}+p_{b})(\rho_{b}+p^{\rm
      def}_{u})}}.
\label{vfldef}
\end{equation}
For $v^{\rm def}_{\rm fl}\ll 1$, one can use (\ref{xieq}) to estimate
the magnitude of the fluid velocity at the shock \cite{Kurki-Suonio85a},
\begin{equation}
v_{\rm fl}^{\rm sh}\lesssim
\exp\left\{-\frac{\left(\frac{1}{v_{\rm def}^{2}}-3\right)}{4\sqrt{3}v_{\rm
    fl}^{\rm def}}\right\}
\label{vflshest}
\end{equation}
To get an idea of the magnitude of $v_{\rm fl}^{\rm sh}$, consider the
values $v_{\rm fl}^{\rm def}=3\times 10^{-3}$ and $v_{\rm def}=0.5$.
(Typically, since $\Tu^{\rm def}-\Tb \ll \Tb$ for the EW phase
transition, $v_{\rm fl}^{\rm def}\lesssim$ a few $\times 10^{-3}$.)
Then we find that $v_{\rm fl}^{\rm sh}\lesssim 10^{-21}$! We can see
that the shock front is {\em extremely} weak. This is due both to the fact
that the phase transition itself is weak (i.e., the latent heat is
small) and that the spherical geometry also drastically `damps' the
strength of the shock wave.

The extreme weakness of the shock front is a generic characteristic of
deflagration bubbles in the EW phase transition. There is an advantage
to this characteristic: it will allow us to solve for $T^{\rm
  def}_{u}$, and ultimately $v_{\rm def}$, by using the following
approximation.

First, consider the velocity of the shock front $v_{\rm sh}$ for a weak
shock. By solving (\ref{rho1p1}) for the shock discontinuity, one
obtains
\begin{equation}
  v_{\rm sh}=\sqrt{\frac{(p^{\rm sh}_{u}-p_{n})(\rho^{\rm
        sh}_{u}+p_{n})}{(\rho^{\rm sh}_{u}-\rho_{n})(\rho_{n}+p^{\rm
        sh}_{u})}}.
\label{vsh}
\end{equation}
Since we have shown that $v_{\rm fl}^{\rm sh}\ll 1$, we can use
(\ref{vflsh}) to estimate $v_{\rm fl}^{\rm sh}\approx\sqrt{3}(\Tu^{\rm
  sh}-\Tn)/\Tn\ll 1$. We can then use (\ref{vsh}) to estimate $v_{\rm
  sh}\approx (1+(\Tu^{\rm sh}-\Tn)/\Tn)/\sqrt{3}$. Therefore, since
$v_{\rm fl}^{\rm sh}$ is so small,
\begin{equation}
v_{\rm sh}\approx \sqrt{1/3}
\end{equation}
to a very good approximation (e.g., one part in $\sim 10^{20}$!).

Now one can employ the conservation of energy to find $\Tu^{\rm def}$.
Consider the volume $V_{0}$ occupied by the entire bubble, including
the shock front. Ignoring the expansion of the universe, which is a
small effect, the total energy inside that volume must be equal to its
initial energy $E_{\rm tot}= \rho_{n}V_{0}$. But $E_{\rm tot}$ must
also be equal to the sum of the $b$ phase part of $V_{0}$ and the $u$
phase part that is shocked. Therefore,
\begin{equation}
\frac{4}{3}\pi v_{\rm sh}^{3}\rho_{n}(\Tn)= \frac{4}{3}\pi v_{\rm
  def}^{3}\rho_{b}(\Tb)+4\pi \int^{v_{\rm sh}}_{v_{\rm
    def}}\rho(\xi)\xi^{2}d\xi.
\label{Ebub}
\end{equation}
Here we have neglected the kinetic energy of the fluid in the shock
because $v_{\rm fl}\ll 1$. Since, for a weak shock, $v_{\rm sh}\approx
\sqrt{1/3}$, the problem is now reduced to finding $\rho_{u}(\xi)$. In
the limit $v_{\rm fl}^{\rm sh}\ll 1$, there is an analytical solution
to (\ref{xieq}) \cite{Kurki-Suonio85a}
\begin{equation}
\rho(\xi)\approx \rho_{u}(\Tu^{\rm
  def})e^{C\left(\frac{1}{\xi}-\frac{1}{v_{\rm def}}\right)}
\label{rho(xi)}
\end{equation}
where $C=v_{\rm fl}^{\rm def}/(1/v_{\rm def}^{2}-3)$.

By setting $v_{\rm def}=v(\Tb,\Tu^{\rm def})$ and combining
equations~(\ref{vdef})(\ref{Ebub})(\ref{rho(xi)}), one can numerically
solve for $\Tu^{\rm def}$ and $v_{\rm def}$. This method vastly
simplifies the problem, because we are just left with essentially two
equations ($v_{\rm def}=v(\Tb,\Tu^{\rm def})$, and (\ref{Ebub})) and
two unknowns ($\Tb, \Tu^{\rm def}$). The reason the problem is
simplified is that the shock is weak and we can set $v_{\rm
  sh}=\sqrt{1/3}$, which eliminates `the third' unknown.

We have compared the results of this approximation with full numerical
calculations (e.i. solving (\ref{vdef}) (\ref{xieq}) etc. directly),
and we have found that the error in temperature $|(T_{\rm app}-T_{\rm
  full})/(T_{\rm full}-\Tc)|<1\%$ for $v_{\rm def}>0.5$ (although as
$v_{\rm def}\rightarrow c_{s}$, this approximation breaks down), and
the error is much smaller for smaller $v_{\rm def}$.  These errors in
temperature correspond to errors $<1\%$ in the calculation of the
deflagration velocity.

\subsection{Numerical results of deflagration wall velocity}

By numerically solving equations~(\ref{vel2}), (\ref{vdef}), and
(\ref{Ebub}), we can calculate the velocity $v_{0}$ of the
deflagration front as a function of the parameters of EW theory ($L$,
$\xi$, $\sigma$, $\Tc$), and the damping coefficient $\eta$.  These
numerical calculations enable us to verify whether our assumption that
the bubble walls travel as deflagrations for a certain region of
parameter space is self consistent, i.e. that $v_{0}<c_{s}$, since for
detonations, $v_{0}>c_{s}$ \cite{Courant85a}. We will also be able to determine
the approximate magnitude of the wall velocity so that we can get an
idea which approximation (see following sections) of the phase
transition is the relevant one.

In figure~3 we have produced some typical bubble temperature profiles
for a couple values of $L$ and $\eta$. Notice that the smaller the damping
$\eta$ is, the higher both $\Tb$ and $\Tu$ are at the bubble wall.
These higher temperatures will in turn prevent the velocity from
becoming very large, because the closer $\Tb$ is to $\Tc$, and the
closer $\Tu$ is to its maximum, the slower the velocity. In this way,
we can see that there is not a simple dependence of $v_{0}$ on
$\eta$. Likewise, larger latent heat will also slow down the bubble
walls because the greater heat released will raise $\Tb$ and $\Tu$.
Notice also that most of the released latent heat is transported in
front of the bubble wall (remember these are spherical bubbles with
$r^{3}$ dependence on volume). We will use this fact later.

Figures~4--7 bear the main fruit of this section:
the deflagration wall velocity as a function of $\eta$ for various
values of the EW parameters. The main point of these graphs is that
there is a wide range of parameter space that produces deflagration
wall fronts.  Recall that we estimated the actual value of $\eta$ to
be of order $\eta \sim g_{W}^{2}\Tc \approx 0.3\Tc$ (see
section~(4.2)).

One the most interesting characteristics of
figures~(4--7) is that $v_{0}$ {\em does not
  increase linearly with} $\eta^{-1}$, as one might naively expect
from equation~(\ref{vel2}). That is to say, the wall velocity depends
not only on the damping coefficient $\eta$, but it also depends on how
quickly heat can be transported away from the wall, because the
pressure difference across the wall is temperature dependent. (Recall
this was briefly discussed in the arguments following
equation~(\ref{damp})).  This process is what one might call the `the
shark-fin effect' (see figure~3): the faster the deflagration wall
goes, the higher the temperature of the plasma around the wall, and
the more the wall resists going even faster.  This effectively acts as
kind of hydro-dynamical damping on the bubble wall, which is in
addition to damping due to internal bubble wall mechanics.
Ultimately, this damping will be limited by heat conduction, which is
not included in these calculations because the transport of heat due
to diffusion is much smaller (in our case) than the transport of heat
from the bulk flows and shock calculated above.

One should always keep in mind that in these calculations, we have
made several assumptions. For example, we have parameterized the wall
thickness with the factor $n_{w}$. As stated before, the above
calculations use $n_{w}=2$ because this is its value when damping and
thermal diffusion are neglected. If anything, one would expect
$n_{w}\gtrsim 2$ because of damping and thermal diffusion. To first
approximation, inspection of (\ref{vel2}) tells us that changing
$n_{w}$ is effectively changing $\eta$. Therefore one can get an
idea of how changing $n_{w}$ will change $v_{0}$ by looking at
figures~4--7 and changing $\eta$ accordingly.

Another approximation was used with the assumed temperature profile of the
bubble wall (3). As stated before, full numerical
calculations show that the shape of $T(x)$ is different than
(3) and there may also be a shift with respect to $\phi(x)$
\cite{Ignatius94a}. These numerical calculations, which only use
conservation of energy and momentum, show that the temperature profile
is shifted toward the $b$ phase. One would also expect that some heat
would thermally diffuse in to the $b$ region because it is at a lower
temperature than the $u$ phase at the wall (see figure~3).  We have
performed calculations including a shift in $T(x)$ with respect to
$\phi(x)$, of the form
\begin{equation}
T(x)= \frac{\Tu+\Tb}{2}+
\frac{\Tu-\Tb}{2}\left(1-2\frac{\phi(x+\epsilon)}{\phi_{b}}\right)
  \label{shift}
\end{equation}
where $\epsilon$ is the shift, and we have found that a shift in the
direction of the $b$ phase ($\epsilon>0$) only {\em decreases} the
velocity of the deflagration front. In fact, a shift on the order of a
correlation length only decreases $v_{0}$ by about $10-20\%$. A shift
in the opposite direction will increase $v_{0}$ ($\epsilon \approx
-\xi$ can increase $v_{0}$ by as much as a factor of two), but the
above mentioned numerical calculations and thermal diffusion make this
shift unrealistic. Therefore we can have some confidence that these
changes will not qualitatively effect our results that are based on
assuming the wall is a deflagration.

\section{Three Scenarios of the phase transition}

The dynamics of the phase transition depend on the velocity $v_{0}$ of
the bubble walls before the the bubbles begin to influence each other.
The behavior of the transition can be separated into three classes:
$v_{0}\geq c_{s}\equiv 1/\sqrt{3}$, $v_{0}\lesssim c_{s}$, and
$v_{0}\ll c_{s}$. Figure~8 presents a qualitative
description of the three cases.

\subsection{Case 1: $v_{0} \geq c_{s}$}

This is the easiest scenario to calculate. Since $v_{0}$ is greater
than the speed of sound, it has been shown that the bubble wall must
propagate as a detonation \cite{Courant85a,Laine94a}. Since no latent
heat is transmitted in front of the bubble wall for detonations (see
figure~8), the bubble walls cannot influence each other until they
actually collide.  Therefore, the velocity of the bubble walls is
unaffected by the presence of other bubbles, and $v(t) = v_{0}$. In
this case there is no baryon enhancement, $\chi=1$, and the baryon
density is homogeneous.

Small inhomogeneities may have been produced at the wall collisions,
but the size of these inhomogeneities $\sim$ bubble wall width. These
inhomogeneities are so small that they would be quickly erased by
thermal diffusion.

\subsection{Case 2: $v_{0} \ll c_{s}$}

When $v_{0}$ is less than the speed of sound, the bubble wall
propagates as a deflagration \cite{Courant85a,Gyulassy84a}. One main
characteristic of a deflagration front is that a shock front precedes
it, with a velocity $v_{\rm sh}>c_{s}$.  This means that some of the
latent heat is transmitted far in front of the deflagration wall, and
this reheated plasma will influence neighboring bubbles. In fact, if
$v_{0}$ is small enough, one can assume that the latent heat released
has been uniformly distributed throughout space. This is the limit of
Case 2: we will assume that all of the latent heat released by the
bubble wall has ample time to completely equilibrate, and the
temperature $T$ is the same everywhere in space.

In this case, since $T=\Tu=\Tb$, we can
use equation~(\ref{velTb}) to define $v(t)$. That is, in the small
velocity limit,
\begin{eqnarray}
  v(t) &=& \frac{n_{w}\xi}{8\eta
\lambda}\left[3B^{2}T(t)^{2}\left(1-\frac{\sqrt{B^{2}T(t)^{2}+8(\lambda
A-B^2)(\Tc^{2}-T(t)^{2})}}{BT(t)}\right)\right.  \nonumber \\
    & & \left.  +12(\lambda
A-B^2)(\Tc^{2}-T(t)^{2})\raisebox{.16in}[.16in][.16in]{\em  } \right].
\label{vsmall}
\end{eqnarray}
Now our problem is reduced to finding $T(t)$, the temperature as a
function of time. This can be done by using the conservation of
energy in the following way.

The pressure $p$ and energy density $\rho$ for the high and low ($u$
and $b$) temperature phases are defined by
\begin{eqnarray}
p_{b}(T)&=& aT^{4} + V(T)\:, \:\:\:\:\:\:\:\:\:\:\:\:\:\:\:\:\:  p_{u}(T)=
aT^{4} \nonumber \\
\rho_{b}(T)&=& 3aT^{4} +T V'(T)-V(T)\:,\:\:\:\:\:\:  \rho_{u}(T)= 3aT^{4}
\label{eos}
\end{eqnarray}
where $a\equiv (\pi^{2}/90)N$, $N=106.75$ is the number of degrees of
freedom of the plasma at $T\simeq 100$GeV \cite{Enqvist91a}, and
$V(T)$ is defined as minimum of $V(\phi,T)$. In order to find $T(t)$,
let us first ignore the expansion of space (we will add it back in
shortly).  Consider a volume $V_{0}$ of space initially in the $u$
phase, and at some temperature $\Tn < \Tc$. Bubbles will be nucleating
and growing in this volume, and the latent heat released will reheat
the plasma in such a way that the total energy in $V_{0}$ will be
conserved:
\begin{equation}
\rho_{u}(T(t))F(t) + \rho_{b}(T(t))(1-F(t))= \rho_{u}(\Tn)
\label{energy}
\end{equation}
If we now take the time derivative of both sides of this equation, it
can be shown, either numerically or by using the simple approximation
\cite{Enqvist91a} $V(T)\simeq L/4(1-T^{4}/\Tc^{4})$, that the second
derivative term $(d^{2}V/dT^{2})$ is negligible, and
equation~(\ref{energy}) becomes
\begin{equation}
\left.\frac{dT}{dt}\right|_{\rm reheat}=
\frac{1}{12aT^{3}}\left(T\frac{\partial
  V(T)}{\partial T}- V(T)\right)\frac{dF(t)}{dt}.
\label{heat}
\end{equation}

The effect of the expansion of the universe is to decrease the energy
density, so we need to add a term to equation~(\ref{heat}) to account
for this. When the universe is not undergoing a change in its equation
of state, the temperature obeys the relation
\begin{equation}
\left.\frac{dT}{dt}\right|_{\rm expansion}= -TH(T)
\label{expand}
\end{equation}
where $H(T)\equiv \sqrt{8\pi\rho(T)/3m^{2}_{\rm pl}}$ is the inverse
Hubble time, and $m_{\rm pl}$ is the plank mass.

Putting eqs~(\ref{heat}) and (\ref{expand}) together, we obtain a
formula for the evolution of the temperature as a function of time:
\begin{equation}
\frac{dT}{dt}= \frac{1}{12aT^{3}}\left(T\frac{\partial V(T)}{\partial
  T}- V(T)\right)\frac{dF(t)}{dt} - T\left(\frac{8\pi(3aT^{4})}{3
  m^{2}_{\rm pl}}\right)^{1/2}.
\label{dT/dt}
\end{equation}
Substituting in the expressions for $F(t)$, $V(T)$, and $v(T(t))$, we
find that (\ref{dT/dt}) is an integro-differential equation which must
be solved numerically. Once $T(t)$ is known, however, one can easily
calculate $v(t)$, and the baryon enhancement factor $\chi$.

The amplitude and size of baryon density inhomogeneities can also be
calculated. As the bubbles grow they will heat up the plasma and slow
down. Since we know that the baryon density $n_{\rm B}\sim v(t)^{-1}$,
we can calculate the baryon density profile the bubble wall leaves
behind as it propagates through space. That is to say, we can
calculate the baryon over density $\delta_{\rm B}(r)$ defined as
\begin{equation}
\delta_{\rm B}(r)\equiv \frac{n_{\rm B}(r)}{n_{\rm B0}}=\frac{v_{0}}{v(r)},
\label{delta}
\end{equation}
where $n_{\rm B0}$ is the baryon density assuming the wall velocity is
constant ($v=v_{0}$), and knowing $v(t)$ at all times allows us to
find the velocity as a function of bubble radius $r$.

One must be careful, however, in using (\ref{delta}), because the
bubbles are nucleating randomly in space, and so the distance a bubble
wall propagates before colliding with another is not well defined. We
can get a good idea of the general behavior and scale by using the
average bubble spacing
\begin{equation}
d(t)=(n_{\rm bubl})^{-1/3}= \left(\int_{t_{c}}^{t}\Gamma_{\rm
nuc}dt\right)^{-1/3}.
\label{d}
\end{equation}
Therefore, $d$ is the scale size for inhomogeneities, and $\delta_{\rm
  B}(r)$ only has meaning out to scales of this order. For the EW
phase transition, one typically finds that $d\sim 10^{-5}H^{-1}$.

Before moving on to the next case, let us make two remarks about this
case. The first remark is actually a helpful rule of thumb. We have
stated before that as the phase transition proceeds, the released
latent heat will reheat the plasma and the velocity of the walls will slow
down. In order to get an idea of how important the reheating of the
phase transition is, we can compare the latent heat $L$ to the energy
needed to bring the plasma back up to $\Tc$ from $\Tn$. If
\begin{equation}
L< \rho(\Tc)-\rho(\Tn)= 3a(\Tc^{4}-\Tn^{4}),
\label{Lsmall}
\end{equation}
then the latent heat will be too small to reheat the temperature back
up to $\Tc$. In this case, we would then expect that the bubble wall
velocity will not slow down very much, and both $\chi$ and
$\delta_{\rm B}$ will be of order unity.

On the other hand, if
\begin{equation}
L\gtrsim 3a(\Tc^{4}-\Tn^{4}),
\label{Lbig}
\end{equation}
then the latent heat will reheat the plasma back up to $\Tc$ and the
velocity of the bubble walls will slow down considerably. In this
case, $\chi\gg 1$, and $\delta_{\rm B}\gg 1$.

The second remark has to do with the case of large latent heat. If the
plasma heats up enough, then then phase transition will almost stop.
It will not stop completely, however, because the expansion of the
universe is continuously removing energy from the plasma, and so the
temperature never quite reaches $\Tc$.  Instead, the temperature
remains constant while the released latent heat goes into expanding
the universe. We can use this fact to estimate the minimum velocity of
the bubble wall by setting $dT/dt=0$ in equation~(\ref{dT/dt}). If we
now use the estimates $F\approx 0.5$, $TV'(T)-V(T)\approx L$,
$\dot{F}/F=\dot{V}/V\approx 3v/r$, where $r$ is the radius of an average
bubble, and $r\approx d/2$, we obtain an estimate for the minimum velocity
\begin{equation}
v_{\rm min}\approx \frac{4dH}{(L/aT^{4})}.
\label{vmin}
\end{equation}
For the EW phase transition, we will find that $v_{\rm min}/v_{0}\sim
1-10^{-3}$, which yields over densities of $\delta^{\rm max}_{\rm
  B}\sim 1-10^{3}$.

\subsection{Case 3: $v_{0} \lesssim c_{s}$}

This is the most difficult case. Since $v_{0} \lesssim c_{s}$ (but
always $<c_{s}$), the bubble walls also propagate as deflagrations, as
in Case~2, and, insofar as the deflagration picture is correct, this
is also the most accurate description of the dynamics of the phase
transition.

In this case a shock front also precedes the deflagration front, but
now the deflagration front is moving too quickly for the released
latent heat to equilibrate with the rest of the plasma. Now there is a
build-up of heat around the deflagration front and there is a
temperature difference across it. The calculation becomes much more
difficult because of the inhomogeneous distribution of temperature,
and the random distribution of bubbles. A numerical simulation is the
only way to solve this case.

One can, however, glean important information from this case without
doing a full numerical simulation. For example, we will use the
deflagration front scenario to carefully calculate the bubble wall
velocity as a function of the damping coefficient $\eta$
(uninfluenced by other bubbles).  One can then determine whether the
deflagration scenario is self-consistent (i.e. $v_{0}<c_{s}$), and one
can also determine how accurate it is to assume the limit of Case~2.
Once we have an accurate picture of uninfluenced bubble wall
propagation, we can then make some rough approximations of the baryon
enhancement factor $\chi$, and on the size and amplitude of the baryon
inhomogeneities produced.

\subsection{Estimating baryon production in case 3:
$v_{0}\lesssim c_{s}$}

As stated before, the complexity of the phase transition prohibits any
accurate calculations without a full numerical simulation. For
example, in this case we have bubbles with shocks and complicated
temperature profiles colliding with each other. Kajantie and
Kurki-Suonio \cite{Kajantie86a} and Ignatius et al. \cite{Ignatius94a}
have studied collisions of deflagration fronts and shock fronts in
(1+1) dimensions, but in (3+1) dimensions, approximations similar to
Case~2 must be made if numerical simulations are not used (e.g. see
\cite{Kurki-Suonio88a}).

Since we are not using a numerical simulation, we will forgo any
detailed calculations and make some general arguments that will give
us a good idea of what is happening during the phase transition in
Case~3. We will also be able to estimate the baryon enhancement $\chi$
and the over density $\delta_{\rm B}$. We will find that Case~3 is very
similar to Case~2.

The first idea to keep in mind is that case 2, which we can calculate,
is a limit of case 3. Therefore, one would expect that the behavior of
the two cases should have some similarities. For example, in case 2
the bubble walls are continuously slowing down due to the reheating of
the plasma as the growing bubbles homogeneously release their latent
heat.  The bubble walls in case 3 should also slow down as the phase
transition proceeds, because the bubble walls are being influenced by
the heated plasma of the shocks of neighboring bubbles.

Let us now describe the general characteristics of the phase
transition for Case~3. First of all, because the shock is so weak in
the EW phase transition, the initial collision of the shock front with
the deflagration front will only result in a slight deceleration of
the deflagration front. It will not stop the wall, at least not at
first. Once the shock has passed through the deflagration front (and
heated it slightly), the hotter plasma behind the shock front will
begin to influence the deflagration front. The temperature of the
plasma behind the shock increases as the two neighboring bubble walls
approach, and this will further slow down the deflagration front.

We can get an idea of how much a bubble wall slows down by considering
the following.  One generic characteristic of deflagration wall
propagation is that most of the latent heat released is transported
{\em in front} of the deflagration wall. Some of the latent heat goes
towards heating the $u$ phase, but most of the latent heat goes in
front of the wall (this is born out in numerical simulations of the
previous section). Therefore, in Case~3, {\em the released latent heat
  is concentrated} into a smaller volume of space, namely the space
immediately in front of the deflagration wall. This is in contrast to
Case~2, where the released latent heat is smoothly distributed
throughout all of space.

One major consequence of this concentration of released latent heat is
that the plasma can now more easily reheat back up to $\Tc$ in these
concentrated regions, so the lower limit for $L$ to produce
significant inhomogeneities found in Case~2 can now be reduced. That
is to say, in Case~3, a larger range of values for $L$ will produce baryon
over densities greater than order unity .

To get an idea of which values of $L$ will produce significant
inhomogeneities, consider the following simple model.
Assume that all the bubbles nucleated at the same time, with their
centers (densely packed) at a distance $d$ apart (see
equation~(\ref{d})). The deflagration walls will travel a distance
\begin{equation}
r_{0}= \frac{v_{0}d}{v_{0}+c_{s}}
\label{r0}
\end{equation}
before encountering a shock, and the velocity of the walls will be
constant ($v_{0}$) out to this distance (recall $c_{s}$ is the velocity
of the shock). At the moment when the deflagration wall encounters
the shock from a neighboring bubble, the volume $V_{u}$ of the unit
cell that is in front of the deflagration front is (therefore still in
the unbroken $u$ phase)
\begin{equation}
V_{u}=
\frac{4\pi
d^{3}}{3(c_{s}+v_{0})^{3}}\left(\left(\frac{c_{s}+v_{0}}{2}\right)^{3}-v_{0}^{3}\right).
\label{Vu}
\end{equation}

If we assume that all of the latent heat is transported in front of the
deflagration wall, then we can estimate how much latent heat is needed
to reheat the plasma in front of the deflagration front back up to
$\Tc$. The total amount of energy that will go into reheating the
plasma in front of the deflagration wall is $E_{\rm tot}\approx
L(V_{u} +4\pi r_{0}^{3}/3)$. Therefore if
\begin{equation}
L\left(1+\frac{v_{0}^{3}}{\left(\frac{c_{s}+v_{0}}{2}\right)^{3}-v_{0}^{3}}\right)\gtrsim 3a(\Tc^{4}-\Tn^{4}),
\label{Lbig3}
\end{equation}
then the latent heat $L$ is big enough to reheat the plasma such that
the bubble walls will slow down considerably once they have collided
with the shocks. This condition must be met in order for $\delta_{\rm
  B}\gg 1$. The only difference between Equation~(\ref{Lbig3}) and the
limit obtained in Case~2 (equation~(\ref{Lbig})) is the factor in the
parenthesis on the left hand side of (\ref{Lbig3}). This is the factor
describing the concentration of the released latent heat. For low
values of $v_{0}$ this factor is close to one. As an example, for
$v_{0}=0.2$, this factor $\simeq 1.16$. But for, say,  $v_{0} =0.5$, this
factor $\simeq 5$. For values of $v_{0}>0.5$, the factor becomes
very large. This means that for large deflagration velocities, the
latent heat $L$ has to be very small in order for the the bubble walls
 to collide without slowing down first.

The scale size $\ell$ for the inhomogeneities in this case is
\begin{equation}
\ell\approx \frac{d}{2}-r_{0}= \frac{d}{2}\frac{(c_{s}-v_{0})}{(c_{s}+v_{0})}
\label{ell}
\end{equation}
for $v_{0}\rightarrow c_{s}$, this can be very small (recall $d\sim
10^{-5}H^{-1}$ is already small).

Now we can put together a picture of what happens during the phase
transition in Case~3. At first the nucleated bubbles travel with
constant velocity $v_{0}$ until they reach a radius $r_{0}$. At this
point they collide with the shock of a neighboring bubble and begin
to slow down. If the latent heat is large enough that condition
(\ref{Lbig3}) is met, the bubble walls will slow down to some minimum
velocity $v_{\rm min}$ (see below), and large amplitude
inhomogeneities of size $\ell$ will form. If the latent heat is small,
then inhomogeneities with amplitude ${\cal O}(1)$ and with size
$\ell$ will form.

The minimum velocity of the deflagration walls can be found by using a
method similar the one used in Case~2 (equation~(\ref{vmin})). First,
let us assume that $L$ is large such that condition (\ref{Lbig3}) is
met. Next, let us make the assumption that the phase transition has
proceeded to the point that the $b$ phase dominates, and the remaining
$u$ phase is in the form of shrinking bubbles of radius $r_{u}\approx
\ell$. The bubbles of the $u$ phase have become so hot that the
deflagration wall has slowed down almost to a stop. In fact, they
would stop were it not for the fact that heat is being removed both by
expansion of the universe and by hydrodynamic leakage (i.e.  shocks
are leaking out of bubble and taking heat with them). Let us make a
gross simplification and assume that the amount of heat leaking out
from the shocks is equal to the amount of heat coming in from shocks
of neighboring bubbles. Then the heat is only removed by the expansion
of the universe, just as in Case~2. We can then estimate the minimum
velocity by using the conservation of energy equation (\ref{dT/dt}).
In this case, we use the same assumptions that led to (\ref{vmin}),
but now $r\rightarrow r_{u}$ and $F$ is the fraction of the bubble
volume that remains after starting with initial radius $\ell$ (we set
$F=0.5$).  With these assumptions, one obtains on estimate for the
minimum velocity of the deflagration wall of the shrinking bubbles,
\begin{equation}
v_{\rm min}\approx
\frac{8r_{u}H}{(L/a\Tc^{4})}\approx\frac{4dH}{(L/a\Tc^{4})}\frac{(c_{s}-v_{0})}{(c_{s}+v_{0})},
\label{vmin3}
\end{equation}
where we have set $r_{u}\approx \ell$. This expression is similar to
Case~2, in fact (\ref{vmin3}) approaches the Case~2 value as
$v_{0}\rightarrow 0$. The only difference between the minimum
velocities for the two cases is the factor
$(c_{s}-v_{0})/(c_{s}+v_{0})$, which accounts for the small size of
the $u$ phase bubbles. For the EW phase transition $v_{\rm min}\sim$
1--$10^{-3}\times(c_{s}-v_{0})$. If $c_{s}\approx v_{0}$ then $v_{\rm
  min}$ can become very small.  However, as stated in section() baryon
production turns off at $v\lesssim 10^{-4}$, so $\delta_{B}^{\rm max}$
cannot become too large.

For the case of large $L$, that is when (\ref{Lbig3}) is met, we can
now estimate $\chi$ and $\delta_{\rm B}$. Estimating the baryon
over density $\delta_{\rm B}$ is simple. Using (\ref{delta}) and
(\ref{vmin3}), one obtains
\begin{equation}
\delta_{\rm B}^{max}\approx
\frac{(L/a\Tc^{4})}{4dH}\frac{v_{0}(c_{s}+v_{0})}{(c_{s}-v_{0})}.
\label{deltamax3}
\end{equation}

The baryon enhancement factor $\chi$ can be estimated by using
\begin{equation}
  \chi \approx \frac{((d/2)^{3}- \ell^{3})+\delta_{\rm B}^{\rm
      max}\ell^{3}}{(d/2)^{3}}
\label{chi3_0}
\end{equation}
where we have used (\ref{nB}) for the density of baryons. Substituting
in equations~(\ref{deltamax3}) and (\ref{r0}) one obtains
\begin{equation}
\chi \approx 1+\left(\frac{c_{s}-v_{0}}{c_{s}+v_{0}}\right)^{3}\left(
\frac{(L/a\Tc^{4})}{4dH}\frac{v_{0}(c_{s}+v_{0})}{c_{s}-v_{0}}-1\right).
\label{chi3}
\end{equation}
As a reminder, equation~(\ref{chi3}) is an estimate of the baryon
enhancement for Case~3 assuming that condition (\ref{Lbig3}) is met,
so that the the deflagration walls slow down to a minimum determined
by the expansion rate of the universe. There is one subtle difficulty
in calculating eq.~(\ref{chi3}), and that is in finding the average
bubble spacing $d$. In the next section we will discuss the behavior
of $d$ in the case of homogeneous heating (Case~2), and we will find
that although $d$ is a function of time, it has an asymptotic value.
For the same reasons as in Case~2 (namely reheating) Case~3 will also
have an asymptotic value of $d$. In order to simplify matters, we will
then use this asymptotic value for $d$ obtained using Case~2 in
eq.~(\ref{chi3}). The numerical results of our Case~3 estimate of
$\chi$ are included in the next section.

\section{Numerical Results}

In this section, we will present and discuss numerical results of the
EW phase transition for various values of the parameters of EW theory.
The numerical calculations will be based on the limit of Case~2, which
assumes that the latent heat released during the phase transition is
homogeneously (and instantaneously) distributed throughout space. We
will also compare these results to approximations obtained for Case~3.

The numerical calculation is done by evolving the phase transition in
discreet time steps much shorter than the duration of the phase
transition. As stated before, the relevant equation of evolution is
(\ref{dT/dt}).

\subsection{Dynamics of the phase transition}

Figure~9 shows the evolution of the temperature of the
universe, velocity of the bubble wall, and the fraction of space
converted to the low $T$ phase during the transition for two
values of latent heat $L$. For large values of latent heat, one
can see that the phase transition quickly reheats the universe close
to $\Tc$, then the phase transition proceeds only because the universe
is expanding and removing the latent heat released.

A larger latent heat actually effects the phase transition in two
ways. First, the larger the $L$, the more heat is released, and
second, a larger $L$ changes the shape of the potential $V(\phi ,T)$
such that bubbles nucleate sooner. This decreases the temperature
difference $\Tc-\Tn$ and makes any released latent heat that much more
effective at reheating the plasma back up to $\Tc$.

As stated previously (sec.~(5.2)), the size of the fraction
$L/3a(\Tc^{4}-\Tn^{4})$ is what determines whether the phase
transition will be appreciably affected by reheating. If the fraction
is of order unity or greater, reheating will play a dominant role in
the evolution of the phase transition. Figure~10 shows this
fraction as a function of $L$ for various values of $\sigma$. For
values of this fraction of order unity or greater, the phase
transitions behaves like the upper curves in figure~9. For
smaller values of this fraction, the phase transition proceeds like
the lower curves of figure~9.

\subsection{Baryon enhancement}

Figure~11 shows the baryon enhancement $\chi$ as a function of $L$,
for different values of $\sigma$, assuming that $n_{B}\propto
v_{0}^{-1}$, as predicted by charge transport mechanism (see
section~2).  Figure~11 is a calculation of $\chi$ for Case~2, the
case where the bubble wall velocity ($v_{0}\ll c_{s}$), and uniform
reheating of the plasma is assumed. Our calculations show that for
Case~2, $\chi$ is relatively insensitive to the initial bubble wall
velocity. This also translates into the fact that for Case~2, $\chi$
is relatively insensitive to the damping parameter $\eta$ (of course,
$\eta$ must be large enough in order for Case~2 to be a valid approximation).

For Case~3, however, our estimate of $\chi$ can be a very sensitive
function of the bubble wall velocity see eq.~(\ref{chi3})), and so it
can also be sensitive to $\eta$. In figure ~12 we have
  plotted $\chi$ as a function of $\eta$ for several values of latent
  heat $L$, in the limit of Case~3. Notice, however, that as $L$
  increases, $\chi$ becomes less sensitive to $\eta$. As mentioned
  before (in the discussion after eq.~(\ref{chi3})), one must be
  careful in choosing the value $d$ of the spacing between bubbles
  when calculating the Case~3 estimate of $\chi$. For figure~12 we
  have used the estimate that $d$ is equal to the asymptotic value
  obtained from case two. See the next section for a complete
  discussion of how $d$ is obtained.

It is interesting to note in figure~12 that $\chi$ has a {\em
  maximum} value as $\eta$ is varied. This can be explained by the
fact that there are two competing effects: as the velocity increases
from zero, (i.e. damping increases), the size $\ell$ of the over dense
regions becomes smaller, decreasing the average baryon density, but at
the same time as the velocity increases, the minimum velocity of the
bubble walls decreases, thus raising baryon production. At some point
between small and large velocities, the combination of the two effects
produces a maximum.

By comparing figure~11 and figure~12, we can
compare the results of baryon enhancement $\chi$ for Case~2 and
Case~3. First, let us recall that the Case~3 estimate is only valid
when the latent heat is large enough that eq.~(\ref{Lbig3}) is
satisfied. In figure~12, this constraint is satisfied for
all values of $\eta$ for the $L^{*}=0.016$ and $L^{*}=0.008$ curves,
but for the $L^{*}=0.004$ curve, the approximation is only valid for
values a $\eta \lesssim 10$GeV.  The values of $\chi$ for both cases
are very similar (within $\sim 30\%$ of each other) for the two larger
values of latent heat.

There are two important observations to make from the comparison of
these two figures. First, for small velocities (i.e. large damping),
where Case~2 is expected to be a good approximation, the two estimates
of $\chi$ roughly agree. Second, for the difficult to (exactly)
calculate scenario of larger bubble wall velocities, where
inhomogeneous heating of the plasma becomes appreciable and Case~3
becomes the valid estimate, we notice that for larger velocities
(smaller damping), $\chi$ decreases, though only by a factor of order
unity. Therefore, instead of having to make the very complicated
calculation of inhomogeneous heating, we estimate that the calculation
of $\chi$ assuming uniform heating of the plasma is a good
approximation, and the effect of inhomogeneous heating only changes
the value of $\chi$ by a factor of order unity.

{}From these two figures, we can conclude that for baryon production
inversely proportional to the wall velocity, there is a wide range of
parameter space that produces a large $\chi > 10$.  Recall that a
large $\chi$ relaxes constraints on other parameters such as $CP$
violation.

Finally, notice that $\chi$ can be a sensitive function of both $L$
and $\sigma$ for certain regions of parameter space. The one-loop EW
value $L_{0}/3a\Tc^{4}\approx 0.004$ is within this sensitive region.

\subsection{Inhomogeneities}

For Case~2 and Case~3, which describe expanding deflagration bubbles,
we have found that the bubble wall velocity decreases as the phase
transition proceeds. This is a generic feature of Case~2 and Case~3.
If one now assumes that baryon production is function of bubble wall
velocity, then inhomogeneities must develop during the phase
transition.

In order to get an idea of the size and amplitude of the inhomogeneities
produced, one can use the calculation of $v(t)$ obtained in the previous
section on the dynamics of the phase transition to determine the
velocity of the bubble wall as a function of bubble wall radius, i.e.
$v(r)$. Then one can apply a velocity dependent theory of baryon
generation to obtain the baryon density as a function of radius from
the bubble center $n_{B}(r)$.

We must be careful, however in interpreting the function $v(r)$,
because our calculation of $v(r)$ assumes that the bubbles expand
without colliding into other bubble deflagration walls. In reality
deflagration walls will collide on many scales due not only to the
fact that the bubble centers are scattered randomly throughout space,
but also because bubbles are continually nucleating during the phase
transition. One can, however, sensibly talk about an {\em average}
spacing between bubbles $d(T)$ (see equation~(\ref{d})), and we can
then assume that the bubble walls propagate a distance $d(T)/2$ before
colliding with other deflagration walls.  This distance decreases with
time, but, as shown in figure~13, $d(T)$ reaches an asymptotic value
during the phase transition. This is because at some point during the
phase transition, the released latent heat has reheated the plasma to
a high enough temperature that nucleation has turned off, and so the
distance between bubble centers remains constant (ignoring the very
slow expansion of the universe).

This asymptotic value for the distance between bubble centers $d_{0}$
sets the scale for the maximum size for the inhomogeneities.
Figure~14 shows $d_{0}$ as a function of latent heat for various
values of $\sigma$. We have found that since $d(T)$ is such a
sensitive function of temperature, it reaches its asymptotic value as
soon as the temperature begins to rise, which is very early in the
phase transition (compare figures~9 and 13), for large
enough values of latent heat. If the $L$ is so small that the
temperature never rises during the phase transition, then $d(T)$
continues to decrease until the phase transition is complete.

Since the number density of bubbles $\approx d(T)^{-1/3}$, we can see
by inspection of figure~13 that since $d(T)$ changes so rapidly, most
of the bubbles are nucleated just before $d(T)$ reaches its constant
value $d_{0}$ and nucleation turns off due to reheating.  Therefore,
since most of the bubbles are nucleated when the average spacing
between bubble centers is constant ($d_{0}$), we can sensibly talk
about bubbles propagating a distance $d_{0}/2$ before colliding with
other deflagration walls, and we then have a reasonable interpretation
of the calculation of $v(r)$. In figure~15, we plotted $v(r)$ for a
bubble nucleated at the peak nucleation time, i.e. the time when the
bubble nucleation rate is the largest, or put another way, the time at
which $\partial n_{\rm bubl}(t)/\partial t =0$, where $n_{\rm bubbl}$
is the density of bubbles. For a large range of parameters we found
that numerically, the peak nucleation time is at most a few$\times
10^{-5}H^{-1}$ before $d(T)$ reaches its constant value.

Not only the size, but also the amplitude of the inhomogeneities
produced can also be inferred from $v(r)$ (figure~15). One can see
that the velocity of the bubble walls can decrease anywhere from a
factor of order unity to a few orders of magnitude, depending on the
values of parameters such as latent heat $L$ or surface tension
$\sigma$. In order to get a more clear idea of the range of
velocities, we calculate the ratio $v_{\rm min}/v_{0}$, where $v_{\rm
  min}$ is the minimum velocity the wall acquires during the phase
transition. We present the result of $v_{\rm min}/v_{0}$ as a function
of latent heat $L$ for various values of $\sigma$ in figure~16. If
the amplitude of inhomogeneities $\delta_{B}\sim v^{-1}$, (see
eq.(\ref{delta})), then figure~16 is also a graph of
$(\delta_{B}^{\rm max})^{-1}$.

In order to determine the density profile for the case when baryon
production is inversely proportional to the bubble wall velocity, as
the charge transport mechanism predicts (see section~2), we have
plotted $v_{0}/v(r)$ in figure~17 (also nucleated at peak nucleation
time as defined above) for two different values of latent heat, where
$r$ is the distance from the bubble center, and $v_{0}$ is the bubble
wall velocity before other bubbles effect the velocity. On this graph,
we have also plotted the average baryon density. Note that for this
case, the inhomogeneities are characterized as large regions of baryon
density which is slightly more dense than average (by a factor of at
most a few), along with large {\em holes} of very low density of
baryons (smaller than the average by up to a few orders of magnitude).
This type of baryon density profile, i.e. one with holes of almost
zero density, is the opposite of the type of density profiles normally
considered for inhomogeneities in the early universe, namely small
regions with large baryon over density surrounded by regions of baryon
density close to the average. We will talk more about these different
types of profiles in the next section.

So far in this section we have only talked about inhomogeneities
generated for Case~2, where the heating is homogeneous. For Case~1
one, as we have discussed before, no significant inhomogeneities are
expected to form at all. But what about Case~3, the most realistic
deflagration wall scenario? As stated in the previous section Case~3
evolves in a way very very similar to Case~2. The inhomogeneities
generated should also be similar. The only difference between Case~2
and Case~3 is that for Case~3, the size of the inhomogeneities will be
smaller and there amplitude larger.

The size of the inhomogeneities for Case~3 will be $\ell$ (see
eq.~(\ref{ell})), which always smaller than $d_{0}/2$, the size of
inhomogeneities for Case~2. We should make it clear here that the {\em
  separation} between the inhomogeneities will be the average bubble
spacing $\approx d_{0}$ for both cases, though the physical {\em size}
of the inhomogeneities is different for each case. Since the minimum
velocity attained is smaller in Case~3, one might expect that the
amplitude of the inhomogeneities for Case~3 will be larger than in
Case~2 (see (\ref{vmin3})and eq.~(\ref{deltamax3}). How much smaller in
size and larger in amplitude the inhomogeneities will be for Case~3
compared to Case~2 depends on how close the bubble wall velocity is to
the speed of sound $c_{s}$ (because the closer to $c_{s}$, the more
concentrated is the released latent heat). For the one-loop EW
parameters (and assuming $\eta=.3\Tc$), for example, $v_{\rm
  min}\approx 0.14$, and the difference between Case~2 and Case~3 is
that the amplitude of the inhomogeneities is only 1.6 times greater
for Case~3 and the size is 1.6 times smaller, assuming that baryon
production $\sim v_{0}^{-1}$.

One should keep in mind, however, that phase transition dynamics and
baryogenesis are {\em separate} issues. We have shown, via phase
transition dynamics, that the velocity of the bubble wall can vary by
as much as a few orders of magnitude. We stress that how this
translates into inhomogeneous baryon production depends on the model
of baryogenesis.  For weak dependence of velocity on baryon
production, this several magnitude variation in $v_{0}$ would only
produce small amplitude inhomogeneities. However, as shown in section
two, there could be {\em very} strong dependence (exponential) of
velocity on baryon production, and this could produce very large
amplitude inhomogeneities. The $v_{0}^{-1}$ dependence that we have
chosen to use as the main example in this paper produces moderately
large amplitude  inhomogeneities.

\section{Effects on primordial nucleosynthesis}

Now that we have determined the size and amplitude of the
inhomogeneities generated in the phase transition for various
scenarios, let us examine whether these inhomogeneities will have any
observable effects on the synthesis of elements in the early universe.
We will see that the crucial factor that will determine whether the
inhomogeneities have any effect on the synthesis of elements will be
their ability to survive the homogenizing effects of diffusion. But
first, let us take a quick look at the theory of element formation in
the early universe. For a review of this topic, see reference
\cite{Walker91a}.

\subsection{Primordial nucleosynthesis}

The synthesis of the elements in the early universe, called
primordial, or Big Bang, Nucleosynthesis (BBN), occurs when the
temperature of the universe has cooled down to the point that neutrons
and protons have a low enough energy to bind together when they
collide.  The abundances of elements, such as Helium, that form depend
on parameters such as the neutron to proton ratio, the baryon density,
and the temperature. The standard BBN calculation (e.g. see
\cite{Walker91a}) assumes that the universe was homogeneous at the
time of nucleosynthesis, and this calculation predicts element
abundances that are consistent with observational constraints on
primordial abundances, although there are still large observational
uncertainties for some of the elements.

There are also BBN calculations that include the presence of
inhomogeneities during nucleosynthesis (for example see
\cite{Fuller88a,Jedamzik94b}), and these calculations produce
abundances significantly different than the homogeneous case. The fact
that inhomogeneities effect the abundances can be understood by
observing that nucleosynthesis reaction rates are a non-linear
functions of baryon density. For example, the more neutrons and
protons there are in a volume, the more likely they are to fuse and
form deuterium, which in turn will quickly join with another deuterium
nucleus to form helium. Specifically, the reaction rate $R$ is
proportional to the product of proton density $n_{p}$ and neutron
density $n_{n}$, $R\propto n_{p}n_{n}=\kappa_{n}n_{p}^{2}$, where
$\kappa_{n}\equiv n_{n}/n_{p}$.  Therefore the reaction rate $R$ is not
linear in density, and inhomogeneities will produce abundances
different than the homogeneous case, even though the average density
(which is a linear function of density) is the same for both cases.

The important result of these inhomogeneous BBN calculations is that
there is a large range of sizes and amplitudes of inhomogeneities that
can be {\em ruled out} because they produce abundances that do not
agree with observation \cite{Jedamzik94b}. So the important question is: Does
any region of EW parameter space produce inhomogeneities that are
ruled out by observations of primordial element abundances?

In order to answer this question, we must first determine what these
inhomogeneities, which were formed early on when the temperature of
the universe was $\sim 100$GeV, look like much later, when
nucleosynthesis takes place, at a temperature $\sim 100$KeV.

\subsection{The evolution of the inhomogeneities up to the
  nucleosynthesis epoch}

Once the inhomogeneities are formed in the plasma, they will
immediately begin to dissipate via diffusion. The time scale for
complete dissipation of the inhomogeneity will depend upon the size
and amplitude of the inhomogeneity, and it will also depend upon the
mean free path of the particles responsible for the diffusion. In the
early universe, there are many different kinds of particles in the
plasma, but it is the particles with the largest mean free path
(though so large that they become decoupled from the plasma) that play
the dominant role in diffusion. For example, for a temperature $T$ in
the range 100GeV$<T<$1MeV, neutrinos have the longest mean free path,
and they are responsible for the dissipation of the inhomogeneities
\cite{Heckler93a,Jedamzik94a}. As the temperature decreases, however,
neutrinos decouple from the plasma, and then the baryons and
eventually photons become the dominant diffusing particles
\cite{Jedamzik94a}.

In order to have an effect on nucleosynthesis, any baryon density
inhomogeneities generated in th EW phase transition must survive until
the nucleosynthesis epoch. If the inhomogeneities are too small in
size (and, strictly speaking, amplitude), they will be dissipated by
diffusive processes, and will have no effect. One can therefore place
a lower limit on the size of inhomogeneities, below which, there are no
observable effects on the abundances of the elements. In order to find
this lower limit, and ultimately if the EW phase transition generates
inhomogeneities larger than this lower limit, let us look closer at
the neutrino, baryon, and photon diffusion processes.

Heckler and Hogan \cite{Heckler93a} and Jedamzik and Fuller
\cite{Jedamzik94a} have studied the dissipation of inhomogeneities due
to neutrino diffusion, and they have found that a wide range of sizes
and amplitudes can easily survive until 1MeV, when other diffusive
processes become important. In particular, for the scales sizes
$d_{0}\approx 10^{-5}H^{-1}(T=100{\rm GeV})\approx10^{-6}$cm. typical
for the EW phase transition, neutrino diffusion will not significantly
effect inhomogeneities with amplitudes smaller than $\sim 10^{4}$, and
amplitudes larger than this would simply decrease until they reached
$10^{4}$. Put another way, neutrino diffusion will have a negligible
affect on the inhomogeneities generated in the EW phase transition.

Therefore the important processes that significantly dissipate the
inhomogeneities generated in the EW phase transition are baryon and
photon diffusion. Jedamzik and Fuller \cite{Jedamzik94a} have shown that for
temperatures well into the nucleosynthesis epoch, baryon diffusion
dominates over photon diffusion. Photon diffusion does play a
significant role in the dissipation of the inhomogeneities for later
times in the nucleosynthesis epoch, but in order to obtain an
(optimistic) estimate of the survival of the inhomogeneities at the
nucleosynthesis epoch, we will neglect photon diffusion and
concentrate on baryon diffusion only.

The important scales for determining the effect of baryon diffusion
are the proton diffusion length and the neutron diffusion length. The
diffusion length of a particle is defined as the average distance the
particle will travel as it random-walks through a plasma for some time
$t$. Here the time $t$ is the time from the initial creation of the
inhomogeneity at the EW phase transition up to the nucleosynthesis
epoch. Because the neutrons have no charge, their mean free path will
be much larger than the proton, and they will have a larger diffusion
length than the proton \cite{Jedamzik94a}. Therefore the neutrons will
diffuse out of baryon over densities much quicker than the protons.

Therefore, since the protons diffuse slower than the neutrons, the
proton diffusion length will be the limiting length scale for the
dissipation of inhomogeneities. That is to say, if the inhomogeneities
are much smaller than the proton diffusion length $d_{p}$, then they
will dissipate before they can have any effect on nucleosynthesis.

Fuller et al \cite{Fuller94a} have calculated the proton diffusion
length integrated from the EW phase transition up to the beginning of
the nucleosynthesis epoch. They have found that for inhomogeneities
with amplitudes less than $10^{2}$, the proton diffusion length is
\begin{equation}
d_{100}^{p}\approx 0.1 {\rm cm}
  \label{dp}
\end{equation}
where $d_{100}^{p}$ stands for the proton diffusion length comoving at 100GeV.
For inhomogeneities with amplitudes larger than $10^{2}$, the comoving
diffusion length decreases. For example, for an amplitude of $10^{4}$,
$d_{100}^{p}\approx 0.01$cm (see figure~18).

How do these values compare with the typical length scale $d_{0}$ for
the inhomogeneities generated in the EW phase transition? As seen from
the previous section, the typical length scales for the
inhomogeneities is $d_{0}\sim 10^{-5}H^{-1}\approx 10^{-6}$cm. In
fact, for a wide range of parameter space
\begin{equation}
\frac{d_{0}}{d_{100}^{p}}\approx 10^{-5}\ll 1.
  \label{dratio}
\end{equation}
Therefore, if inhomogeneities are generated on the scale of $d_{0}$,
and have amplitudes $\lesssim 10^{-4}$, they will all dissipate before
they can effect nucleosynthesis.

Figure~18 presents $d_{100}^{p}$ as a function of amplitude
(obtained directly from Fuller et al \cite{Fuller94a}). Also plotted
on this figure are values of the average spacing between bubble
centers $d_{0}$ (for various values of latent heat $L$ and bubble
surface tension $\sigma$), which is assumed to be the scale size for
inhomogeneities generated in the EW phase transition. The amplitude of
the inhomogeneities for these points was obtained by assuming that
baryon generation is inversely proportional to bubble wall velocity,
as predicted by the charge transport mechanism of baryogenesis. This
figure illustrates very well that $d_{0}$ is much smaller than
$d_{100}^{p}$ for a large range of parameter space.

In order to get a more general idea of the size of the inhomogeneities
produced in the EW phase transition, we have also plotted the mass and
over density of the inhomogeneities in figure~19, and compared these
with the masses and over densities for which proton and neutrino
diffusion is important. Notice that the inhomogeneities generated are
about $10^{-30}$ solar masses in size, so we can see that in terms of
the universe today, these inhomogeneities are very small in scale.

\subsection{Possibilities for generating larger inhomogeneities}

We have seen in the last section that the average spacing between
bubbles $d_{0}$ is much smaller than the comoving proton diffusion
length, and if inhomogeneities are produced at this scale, they will
dissipate before they can effect nucleosynthesis. However, if
inhomogeneities are somehow generated on scales a few orders of
magnitude larger than $d_{0}$ (depending on the amplitude of the
fluctuations), then they would be able to survive until the
nucleosynthesis epoch and effect the abundances of the elements.
Below, we examine a plausible scenario for producing inhomogeneities
on scales much larger than $d_{0}$ \cite{Hogan94a}.

For example, consider the following plausible scenario for producing
inhomogeneities much larger than $d_{0}$. From our observations of
many known phase transitions (such as liquid boiling), we find that it
is very common for phase transitions to be induced by `impurities' in
the system, long before thermal nucleation has a chance to play a
role. Let us apply this idea to the EW phase transition: what if
bubble nucleation in the EW phase transition was induced by some
impurity, or seed, rather than by thermal nucleation?

First of all, if the phase transition was induced by a seed, the
bubbles would be nucleated at a temperature much closer to $\Tc$ than
if it was induced by thermal nucleation. Any latent heat released from
the nucleated bubbles would therefore be much more effective at
reheating the plasma back up to $\Tc$, where nucleation ceases. If the
plasma was reheated quickly back up to $\Tc$ and further bubble
nucleation was turned off, one would expect that the scale size for
inhomogeneities would now be the average bubble spacing at the time
when nucleation turned off, just as in the thermal nucleation case.
But in the seed nucleation case, the scale for average spacing between
bubbles now depends the scale $d_{\rm seed}$ associated with the
density (and efficiency of nucleation) of the seeding agent.

Even though we can only speculate on what the value of $d_{\rm seed}$
could be, we can still place an upper limit on the largest spacing
between bubbles in the seed nucleation scenario, independent of
$d_{\rm seed}$. This can be done by observing that, although it is
true that the seed nucleated bubbles can reheat the universe back up
to $\Tc$, it will still take a finite amount of time to do so, because
the shocks carrying the released latent heat (we are assuming the
bubble walls are deflagrations) can only travel at a speed $v_{\rm
  sh}<c$. If the shocks take too long to heat up the universe, thermal
nucleation will start to become important, and more bubble will start
to nucleate, decreasing the average spacing. The maximum amount of
time $t_{\rm max}$ that the shock waves can have to reheat the plasma
before thermal nucleation becomes important can be estimated as
\begin{equation}
t_{\rm max}\approx t(\Tn)-t(\Tc)
  \label{tmax}
\end{equation}
where $t(T)$ is defined as the time at which the universe is at a
temperature T, and $\Tn$ is the temperature at which the maximum number of
bubbles is thermally nucleated (see section~(6.3)). This time can be
translated into a distance $d_{\rm max}$. If we approximate $v_{rm
  sh}\approx 1$, we find that for a wide range of EW parameter space
\begin{equation}
d_{\rm max}\approx 10^{-3}H^{-1}\approx 10^{-3} {\rm cm}
  \label{dmax}
\end{equation}
where $H^{-1}$ is the Hubble length at $T\approx 100$GeV. Notice that
this is about two orders of magnitude bigger than $d_{0}$. However,
this is still smaller than the proton diffusion length (see
figure~18).

There is still the scale size $d_{\rm seed}$ to consider. For example,
if there is some large scale coherence $d_{\rm seed}$ on the scale of
a Hubble length associated with the seeding agent, then this would
also produce inhomogeneities. Let us use an example. One possible
seeding agent in the EW phase transition is cosmic strings
\cite{Yajnik86a}. If the density of these strings varied on scales
comparable to the Hubble length, then one would expect that the amount
of bubble seeding would also vary on the scale of a Hubble length.
Regions with a large density of strings would nucleate very close to
$\Tc$, and as described above, these regions would also reheat quickly,
and the bubble walls would quickly slow down. On the other hand,
regions with low string density will nucleate at temperatures much
farther away from $\Tc$, and these regions would not reheat so
quickly. The bubble walls in these low density string regions would
then propagate faster, because the temperature is farther away from
$\Tc$. Since the average bubble wall velocities are different in these
two regions, one would expect that the baryon density would be
different, and the scale size for the these regions would be $d_{\rm
  seed}$.

To summarize, it is plausible that the seed nucleation scenario
(caused for example by the presence of cosmic strings) could provide
two scales for producing inhomogeneities: $d_{\rm max}$ and $d_{\rm
  seed}$. We have shown that although $d_{\rm max}$ is larger than the
expected size for inhomogeneities generated via thermal nucleation, it
is still much smaller than the proton diffusion length, and so they
will not survive to effect nucleosynthesis.  However, the seed
nucleation scenario can also produce inhomogeneities on the scale
$d_{\rm seed}$, which can be larger than the proton diffusion length,
though the actual size of $d_{\rm seed}$ is presently still
speculation.

\section{Conclusion}

The main conclusion of this paper is that phase transition dynamics
can play an important role in electro-weak baryogenesis. We have shown
that through the course of the phase transition, the bubble wall
velocity can slow down by as much as a few orders of magnitude,
depending on the values of the presently unknown parameters of the EW
phase transition such as latent heat and bubble wall surface tension.
Since all baryon production mechanisms are sensitive to bubble wall
velocity for at least some range of velocities, the dramatic decrease
in bubble wall velocity can change the average baryon density a few
orders of magnitude from what is expected from calculations which
assume that the bubble wall velocity is constant.  In fact, we
calculated the baryon production in the EW phase transition using the
model called the charge transport mechanism, and we found that the
average baryon density could be as much as $10^{2}$ greater than the
constant bubble wall velocity calculation.  This enhancement of baryon
production could ease the constraints (from the observed average
baryon density) on other parameters of the theory, such as the amount
of required $CP$ violation.

In addition to affecting the average baryon density, the changing
bubble wall velocity also produces inhomogeneities. The amplitude of
the inhomogeneities depends upon the dominant mechanism of
baryogenesis. For the charge transport mechanism, we found amplitudes
as great as $10^{2}$. The physical size of the inhomogeneities is a
relatively insensitive function of the parameters of the EW phase
transition, and we found that the characteristic scale size for
inhomogeneities is $\approx 10^{-5}H^{-1}$, where $H^{-1}$ is the
Hubble length at the phase transition. These inhomogeneities are to
small to survive until the nucleosynthesis epoch and effect the
abundances of the elements. Instead they dissipate via baryon diffusion
before nucleosynthesis begins. There are, however, plausible (though
speculative) scenarios that can produce large enough inhomogeneities
that would effect nucleosynthesis. For example, if bubbles are
nucleated by some seed such as cosmic strings, then the string density
scale may be large enough to create large inhomogeneities.

Besides the topic of baryogenesis, we have also made some interesting
conclusions about bubble wall propagation in the EW phase transition.
Most notably, we have found that for a very wide range of parameter
space, the bubble walls travel as deflagrations. There is not only
`frictional' damping on the wall, which we parameterized with a
damping parameter $\eta$, but we have also found that with the
inclusion of the conservation of energy and momentum, there is an
additional `hydro-dynamical' damping on the wall which is important,
especially when the frictional damping $\eta$ is small. The
consequence of this hydro-dynamical damping is to lower the bubble wall
velocity to a point that all but the most weakly damped walls travel
as deflagrations.  It is important to show that the walls travel as
deflagrations because it is only this type of bubble wall propagation
(i.e. one with a shock that precedes it) that produces the interesting
behavior of dramatic bubble wall deceleration due to reheating.

The author would like to thank Peter Arnold, Larry Yaffe and
especially Craig Hogan for many helpful comments and discussions. This
work was supported by NASA grant NAGW 2569 and NAGW 2523 at the
University of Washington.

\bibliographystyle{prsty}
\bibliography{bubble}

\newpage

{\bf Figure~1:} A schematic picture (not to scale) of the
general dependence of baryon production on bubble wall velocity. For
small velocities, there is an exponential dependence. For velocities
close to the speed of light, baryon production is suppressed because
time scales become too short for baryon violating processes to create
baryons. The velocity dependence in between the dashed line depends on
the mechanism of baryon production. Here we have made a simple
interpolation of what the velocity dependence might look like between
the dashed lines. This shape is very similar to the velocity
dependence of the charge transport mechanism examined in the text.

{\bf Figure~2:} Schematic of fluid velocity and temperature
profiles for spherically expanding deflagration and detonation fronts.
In both cases the fronts are propagating from left to right. Since the
bubble wall quickly reaches a steady state of propagation after
nucleation (see text), to a good approximation the profile is a
function of $r/t$, where $r$ is the distance from the bubble center
and $t$ is the time since nucleation.

{\bf Figure~3:} Temperature profiles for spherical
deflagration bubbles for various values of damping coefficient $\eta$
and latent heat parameter $L'\equiv L/(3aT_{c}^{4})$.  The
discontinuity in the temperature is the deflagration front: it is
where the phase is changing from the unbroken phase to the broken
phase.  In front of the discontinuity is the shock front, where most
of the latent heat is transported. Notice that as the wall goes
faster, more heat ``bunches up'' around the deflagration front.  This
shark-fin effect is responsible for a damping of the deflagration
front in addition to the damping from $\eta$.

{\bf Figure~4:} Deflagration wall velocity $v_{0}$ as a
function of the damping coefficient $\eta$, calculated using
eq.~(4.18). The velocity is calculated with several different values
of the latent heat parameter $L^{*}\equiv L/(3aT_{c}^{4})$. Unless
otherwise stated, in figures~4--7 the parameters
take the value $L^{*}=0.004$, $\sigma =0.005T_{c}^{3}$, $\xi =
15/T_{c}$ and $T_{c} = 100$GeV. These values are close to the one-loop
calculated values, see eq.~(3.3). Some of the lines in the figures
stop. For example, the $L^{*}=0.001$ line stops where $v_{0}\approx
c_{s}$ because our approximation (4.18) breaks down.  One might expect
that the wall becomes a detonation at this point \cite{Ignatius94a}.
Note that EW theory predicts that $\eta \sim g_{W}^{2}T_{c}\sim 30$GeV
(see eq.~(\ref{etaapprox})). For a wide range a parameter space, this
puts $v_{\rm def}$ well below $c_{s}$.

{\bf Figure~5:} Same as figure~4, but with several
values of the bubble wall surface tension $\sigma$.

{\bf Figure~6:} Same as figure~4, but with several
values of the Higgs correlation length $\xi$.

{\bf Figure~7:} Same as figure~4, but with several
values of the critical temperature $T_{c}$.

{\bf Figure~8:} A qualitative picture of the three possible
scenarios of the phase transition. The top three boxes are snapshots
of the phase transition for each case just before the bubble walls
collide. The arrows begin at the wall fronts and indicate which
direction the walls are propagating. The bottom three boxes are
snapshots just moments after the bubble walls have collided. Each box
includes a qualitative temperature and baryon density profile. Notice
that in Case~2 the temperature is homogeneous, and the bubble wall
front is indicated with a dashed line. Note also that since all
temperature fluctuations are much smaller than $T_{c}$, their affect
on baryon density is negligible.

{\bf Figure~9:} The evolution of temperature, fraction $F$
of unbroken phase remaining, and bubble wall velocity during the phase
transition, as a function of time in units of $10^{-5}H^{-1}$, where
$H^{-1}$ is the inverse Hubble time. The solid line is for the case
$L/(3aT_{c}^{4})=0.008$, and the dashed line is for the case
$L/(3aT_{c}^{4})=0.004$. All of the other parameters are given the EW
one-loop values ($\sigma =\sigma_{0}$ etc.). The initial time $t=0$ is
arbitrarily defined as the time when $10^{-6}$ of the universe has
converted to the low temperature phase.

{\bf Figure~10:} The fraction $L/(3a(T_{c}^{4}-T_{n}^4)$ as a
function of $L$ for various values of $\sigma$. This fraction
determines the importance of reheating in the phase transition.  For
values greater than unity, reheating plays an important role, and the
bubble wall velocity will change by a large factor during the course
of the phase transition.

{\bf Figure~11:} Baryon enhancement factor $\chi$ as a
function of latent heat parameter $L'\equiv L/(3aT_{c}^{4})$, for
various values of surface tension $\sigma$, for the case of uniform
heating (Case~2). The quantity $\sigma_{0}$ is the one-loop EW theory
value, defined in the text. All other values of the parameters are
also the one-loop values. Here we use $\Gamma =0.03/$GeV, though
$\chi$ is relatively insensitive to $\Gamma$.

{\bf Figure~12:} Baryon enhancement factor $\chi$ as a
function of the damping coefficient $\eta$ for various values of
latent heat $L$, for the case of non-uniform heating (Case~3). All
other values of the parameters are also the one-loop values.

{\bf Figure~13:} The average spacing between bubble centers as
a function of time for two different values of latent heat $L$.  The
notation and $t=0$ in this figure is the same as in figure~9.

{\bf Figure~14:} The asymptotic value $d_{0}$ of the bubble
center spacing as a function of latent heat, for various values of
$\sigma$.

{\bf Figure~15:} The plot $v(r)/v_{0}$ for
$L/(3aT_{c}^{4})=0.008$ (solid line) and for $L/(3aT_{c}^{4})=0.004$
(dashed line, where $v(r)$ is the bubble wall velocity, $r$ is the
bubble radius, and $v_{0}$ is the initial bubble wall velocity. We
have indicated the radius $d_{0}/2$ at which the bubbles will, on
average, collide. Note that if baryon production is proportional to
the wall velocity, then this is also a plot of the density profile as
a function of distance $r$ from the bubble center.

{\bf Figure~16:} The minimum velocity (divided by the initial
velocity) that a bubble wall acquires during the phase transition for
Case~2 as a function of latent heat $L$, and for various values of
$\sigma$.

{\bf Figure~17:} The plot $v_{0}/v(r)$ for
$L/(3aT_{c}^{4})=0.008$ (solid line) and for $L/(3aT_{c}^{4})=0.004$
(dashed line, where $v(r)$ is the bubble wall velocity, $r$ is the
bubble radius, and $v_{0}$ is the initial bubble wall velocity. For
baryon production proportional to the inverse bubble wall velocity,
this is also the baryon density profile of the bubble. We have also
plotted the baryon enhancement $\chi$ for both cases (e.g., the upper
dash-dot line is for the $L/(3aT_{c}^{4})=0.008$ case) because in
these units normalized by $v_{0}$, $\chi$ is the average baryon
density.

{\bf Figure~18:} The proton diffusion length $d_{100}^{p}$,
comoving at 100GeV, as a function of amplitude of inhomogeneity.
These values of $d_{100}^{p}$ are approximate, and they are taken
directly from Fuller et al \cite{Fuller94a}. We have also plotted
lines which indicate the size $d_{0}$ and amplitude of inhomogeneities
generated in the EW phase transition, for various values of latent
heat $L$ and bubble surface tension $\sigma$.  Each line represents a
line of constant $\sigma$. $\sigma= 2\sigma_{0}$ (long dash), $\sigma=
\sigma_{0}$ (short dash), and $\sigma= 0.5 \sigma_{0}$ (dot-dash). The
lines run from $L/(3aT_{c}^{4})=0.001$ to $L/(3aT_{c}^{4})=0.016$ from
left to right. Notice that for any of these values, $d_{0}\ll
d_{100}^{p}$.

{\bf Figure~19:} The mass scale of inhomogeneities produced
in the EW phase transition (with the same notation and data as in
figure~18), compared with mass and over density scales for
which proton and neutrino diffusion are important. The neutrino
diffusion scales were taken directly from Jedamzik and Fuller
\cite{Jedamzik94a}.

\end{document}